\documentclass[%
 aip,
amsfonts,amsmath,amssymb,
preprint,%
author-year,%
]{revtex4-1}

\usepackage{bm}
\usepackage{color}
\usepackage{graphicx}
\usepackage[T1]{fontenc}
\usepackage[utf8]{inputenc}
\usepackage[mathlines]{lineno}
\usepackage{stackengine,scalerel,stackrel}
\usepackage{fancyhdr}

\newcommand\ie{i.e.\ } 
\newcommand\eg{e.g.\ } 
\newcommand\dd{{\rm d}} 
\def\Qcut{\ThisStyle{\ensurestackMath{\stackon[-2.4\LMpt]{%
\SavedStyle Q}{\kern-.5pt\kern\LMpt\rule{2.5\LMex}{.25pt+.15\LMpt}}}}} 
\begin{document}

\thispagestyle{empty}

\title{\Large{Scattering by source-type flows in disordered media}}

\author{Gerardo SEVERINO$^1$ and Francesco GIANNINO$^2$ \\
\vspace{5pt}
\small{
$^1$\textrm{\textit{Division of Water Resources Management}, University of Naples - \textsc{Federico II}
\\ via Universit\'{a} $100$ - \textsc{I}$80055$, Portici $($NA$)$, \textsc{Italy}}} $($\textrm{e-mail: gerardo.severino@unina.it}$)$ \\
\vspace{5pt}
\small{
$^2$\textrm{\textit{Division of Ecology and System Dynamics}, University of Naples - \textsc{Federico II}
\\ via Universit\'{a} $100$ - \textsc{I}$80055$, Portici $($NA$)$, \textsc{Italy}}} $($\textrm{e-mail: francesco.giannino@unina.it}$)$
}

\date{\today}

\begin{abstract}

Scattering through natural porous formations (by far the most ubiquitous example of disordered media) represents a formidable tool to identify effective flow and transport properties. In particular, we are interested here in the scattering of a passive scalar as determined by a steady velocity field which is generated by a line of singularity. The velocity undergoes to erratic spatial variations, and concurrently the evolution of the scattering is conveniently described within a stochastic framework that regards the conductivity of the hosting medium as a stationary, Gaussian, random field. Unlike the similar one for uniform (in the mean) flow-fields, the problem at stake results much more complex.  \\
\indent Central for the present study is the fluctuation of the driving field, that is computed in closed (analytical) form as large time limit of the same quantity in the unsteady state flow regime. The structure of the second-order moment~$X_{rr}$, quantifying the scattering along the radial direction, is explained by the rapid change of the distance along which the velocities of two fluid particles become uncorrelated. Moreover, two approximate, analytical expressions are shown to be quite accurate into reproducing the full simulations of~$X_{rr}$. \\
\indent Finally, the same problem is encountered in other fields, belonging both to the classical and to the quantum physics. As such, our results lend themselves to be used within a context much wider than that exploited in the present study.
\end{abstract}
\keywords{source-type flow$\, \, \cdot \, \,$scattering$\, \, \cdot \, \,$stochastic modelling$\, \, \cdot \, \,$radial moment}

\maketitle
\thispagestyle{empty}

\section*{Introduction and problem formulation}
Scattering processes represent one of the most powerful diagnostic-tool in applied sciences. In quantum physics, they serve to infer the structure as well as the charge-density of particles~\citep{martin2019}, whereas in the theory of composites they enable one to identify effective flow and transport properties~\citep{dagan1989}. In the classical electrodynamics a similar problem is encountered when one aims at computing the electric field generated by a localized/distributed density of charges~\citep{jackson2007}. \\
\indent In the present study, we are interested in a specific category of scattering phenomena, namely those associated to:
\begin{equation} \label{model}
\mathcal{H} \left( \bm a \right) \equiv a_m \int \dd \bar{\bm{x}} \, \exp \left( \jmath \, \bm a \cdot \bar{\bm{x}} \right) G^\infty_3 \left( \bar x \right) \frac{\partial}{\partial \bar{x}_m} \, G^\infty_2 \left( | \bm{x}_r - \bm{\bar x}_r | \right), \qquad m = 1, 2,
\end{equation}
where~$a_m$ are either real or complex constants, whereas
\begin{equation} \label{GFs}
G^\infty_d \equiv \frac{1}{4 \pi}
\begin{cases}
\ln x_r^{-2} & \quad d = 2 \\
\, \, \, x^{-1} & \quad d = 3
\end{cases} 
\end{equation}
is the $d$-dimensional Green function. Moreover, $\bm{x}_r$ and~$\bm x$ represent the position in~$\mathbb{R}^2$ and $\mathbb{R}^3$, respectively. The explicit evaluation of the~$\mathcal H$-function is crucial for numerous branches of physics. For instance, in nuclear Physics it provides the Born's approximation of the cross section in a Rutherford-type scattering due to the potential:
\begin{equation} \label{density}
\mathcal{U} \equiv \bm a \cdot \nabla G^\infty_2 \left( | \bm{x}_r - \bm{\bar x}_r | \right)  G^\infty_3 \left( \bar x \right)
\end{equation}
\citep[see \eg][]{tanaka2016}. Likewise, in the solid state physics the integral~\eqref{model} accounts for~\textsf{X}-rays scattering~\citep[a general overview can be found in][]{blanco2014} within a quantum system of electrons with density~\eqref{density}. Not disregarded, in electrostatic the quantity~$\mathcal H \left( \bm a \right)$ coincides with the potential generated by the continuous distribution:
\begin{equation}
\rho \equiv \exp \left( \jmath \bm a \cdot \bm x \right) \bm a \cdot \nabla G^\infty_2 \left( x_r \right)
\end{equation}
of electric charge~\citep{renau1982}. Finally, in the reservoir engineering the integral~$\mathcal H$ constitutes the starting point to model scattering  generated by injecting/pumping wells operating in heterogeneous porous formations. Despite the importance of~\eqref{model} for a wide range of applications, hereafter we shall focus on its role in the theory of disordered media. \\
\indent Thus, we consider a steady flow generated by a line-source embedded in a porous formation, and we aim at quantifying the scattering consequent to the injection of a passive scalar. The medium is, as a rule in natural formations, heterogeneous with the conductivity~$K$, in particular, changing erratically in the space by orders of magnitude~\citep{rubin2003}. Such a variability affects tremendously scattering, as it was demonstrated both theoretically~\citep{koplik94,leborgne08} and experimentally~\citep{kurowski1994}. The approach to account for these variations, and to model the associated uncertainty, is to regard the log-conductivity~$\ln K \left( \bm x \right)$ as a stationary, Gaussian, random field. As a a consequence, the dependent flow and transport variables become stochastic, and we wish to characterize scattering by means of the first and second-order spatial moments that, by virtue of ergodicity, are:
\begin{equation} \label{final}
\left \langle \bm X (t) \right \rangle = \bm R \left( t \right), \qquad \qquad \left \langle X^\prime_m (t) X^\prime_n (t) \right \rangle = X_{m,n} (t) \qquad \quad m,n = 1,2,3
\end{equation}
(hereafter~$\langle \rangle$ shall denote the ensemble average operator). Hence, central for the study of scattering are the mean~$\bm R$ and the  fluctuation~$\bm X^\prime = \bm X - \bm R$ of the trajectory~$\bm X \equiv \bm X \left( t \right)$ of a fluid particle. \\
\indent Unlike scattering driven by mean uniform flows~\citep[an exhaustive overview can be found in][]{dagan1989}, here computing the fluctuation~$\bm X^\prime$ is an extremely complex problem~\citep[see, \eg][]{tartakovsky08}, due to the strong coupling of the velocity field~$\bm V$ with the spatial variability of~$K$. A simplification is achieved~\citep[for details, see][]{Indelman1996} by dealing with a weakly heterogeneous formation~\citep[in close analogy to the frozen turbulence approximation,][]{Bahraminasab08}, which leads to the following system of scattering equations:
\begin{equation} \label{trajectory}
\begin{cases}
\dot{\bm R } = \bm U \left( \bm R \right), \qquad \bm R \left( 0 \right) = \bm R_0 , \\
\dot{\bm{X}}^\prime - \bm{X}^\prime \cdot \nabla \, \bm U \left( \bm R \right) = \bm{u} \left( \bm R \right), \qquad \bm{X}^\prime \left( 0 \right) \equiv \left( 0, 0, 0 \right),
\end{cases}
\end{equation}
being~$\bm U \equiv \left \langle \bm V \right \rangle$ and ~$\bm u = \bm V - \bm U$ the mean and the fluctuation of the velocity, respectively. In order to compute the latter, we start from the governing flow equation:
\begin{equation} \label{mp}
- \nabla \cdot \left[ K \left( \bm x \right) \nabla H \left( \bm x \right) \right] = \frac{\bar Q}{\langle K \rangle} \, K \left( 0, 0, x_3 \right) \delta \left( \bm{x}_r \right), \qquad \quad \lim_{x \to \infty} H \left( \bm x \right) = 0
\end{equation}
\citep{severino2020uncertainty}, where the specific energy (the head)~$H \equiv H \left( \bm x \right)$ is related to the velocity~\textit{via} the constitutive model $\bm V = - \left( K / n \right) \, \nabla H \left( \bm x \right)$. The porosity~$n$, in line with the experimental data~\citep{rubin2003}, is regarded as a given constant, whereas~$\bar Q$ is the specific (per unit length) strength. We cast the mathematical problem~\eqref{mp} in dimensionless form by introducing the scaled coordinate~$\bm{x} / \ell_c$, where the characteristic length-scale will be chosen later on. Hence, introduction of the normalized fluctuation $Y \equiv \ln \left( K / K_G \right)$ ($K_G \equiv \exp \left \langle \ln K \right \rangle$ is the geometric mean) transforms eq.~\eqref{mp} (for simplicity we keep the former notations) as follows:
\begin{equation} \label{neweq}
- \nabla^2 H \left( \bm x \right) = Q \, \delta(\bm{x}_r) + \nabla Y \left( \bm x \right) \cdot \nabla H \left( \bm x \right), \qquad \qquad Q \equiv \frac{\bar Q}{\langle K \rangle \, \ell_c},
\end{equation}
where we have accounted for~$K \left( 0, 0, x_3 \right) \delta \left( \bm{x}_r \right) \equiv K \left( \bm x \right) \delta \left( \bm{x}_r \right)$. Solving eq.~\eqref{neweq} is a formidable and quite complex task, with no exact solution. As a matter of fact, one has to sort with approximate methods. In the present paper we adopt a strategy which ultimately leads to simple (analytical) results. More precisely, we expand the head into asymptotic series~$H = H^{(0)} + H^{(1)} + \dots$ of~$Y$ with~$H^{(n)} = \mathcal O \left( Y^n \right)$, and substitute into~\eqref{neweq} to get the governing equations for the leading-order term~$H^{(0)}$ and the fluctuation~$H^{(1)}$:
\begin{equation} \label{flowes}
- \nabla^2 H^{(0)} = Q \, \delta(\bm{x}_r) \Rightarrow H^{(0)} \left( x_r \right) = Q \, G^\infty_2 \left( x_r \right), \, \, \, \, - \nabla^2 H^{(1)} \left( \bm x \right) = \nabla_r H^{(0)} \left( x_r \right) \cdot \nabla_r Y \left( \bm x \right),
\end{equation}
being~$ \nabla_r \equiv \left( \frac{\partial}{\partial x_1}, \frac{\partial}{\partial x_2} \right)$ the gradient in the horizontal plane. Once the second of~\eqref{flowes} is solved, the mean~$U = \langle V \left( x_r \right) \rangle$ and the fluctuation~$\bm u$ of the velocity field are obtained upon expansion of the constitutive model, \ie
\begin{equation} \label{u}
U \left( x_r \right) = \frac{Q K_G}{2 \pi n x_r} \, , \qquad \quad \bm u \left( \bm x \right) = U \left( x_r \right) Y\left( \bm x \right) - \left (\frac{K_G}{n} \right)  \nabla H^{(1)} \left( \bm x \right).
\end{equation}
Hence, the mean~$R$ and the fluctuation~$X^\prime$ of the trajectory are computed by carrying out the quadrature in~\eqref{trajectory} with zero initial condition, \ie
\begin{equation} \label{Xp}
R \left( t \right) = \left( \frac{Qt}{n \pi} \right)^{1/2} \, , \qquad \quad X^\prime \left( R \right) = U \left( R \right) \int^R_0 \dd x_r \, \frac{u \left( x_r, \theta, 0 \right)}{U^2 \left( x_r \right)}
\end{equation}
(we have switched to~$R$ as independent variable, and taken~$\ell_c / K_G$ as characteristic time scale). Moreover, since we are concerned with radial scattering, we have set~$x_3 = 0$. The second-order moment~$X_{rr}$ writes as:
\begin{equation} \label{Xrrr}
X_{rr} \left( R \right) = \left \langle X^{\prime \, 2} \right \rangle = U^2 \left( R \right) \int^R_0 \int^R_0 \dd x^\prime_r \, \dd x^{\prime \prime}_r \, \frac{u_{rr} \left( x^\prime_r, x^{\prime \prime}_r \right)}{U^2 \left( x^\prime_r \right) U^2 \left( x^{\prime \prime}_r \right)}.
\end{equation}
It is worth noting that the covariance $u_{rr} \left( x^\prime_r, x^{\prime \prime}_r \right)  \equiv \left \langle u \left( \bm{x}^\prime_r \right) u \left (\bm{x}^{\prime \prime}_r \right) \right \rangle$ does not depend upon the anomaly~$\theta$, due to the axial symmetry of the mean flow, and it is obtained straightforwardly from the second of~\eqref{u}, the final result being:

\begin{align}
u_{rr} \left( x^\prime_r, x^{\prime \prime}_r \right) = & \, \sigma^2_Y \, \rho_Y \left( \left | \bm{x}^\prime_r - \bm{x}^{\prime \prime}_r \right |\right) U \left( x^\prime_r \right) U \left( x^{\prime \prime}_r \right) + \left( \frac{K_G}{n} \right)^2 \frac{\partial^2}{\partial x^\prime_r \partial x^{\prime \prime}_r} \left \langle H^{(1)} \left( \bm{x}^\prime_r \right) H^{(1)} \left( \bm{x}^{\prime \prime}_r \right) \right \rangle - \notag \\
& \frac{K_G}{n} \left [ U \left( x^\prime_r \right) \frac{\partial}{\partial x^{\prime \prime}_r} \left \langle Y \left( \bm{x}^\prime_r \right) H^{(1)} \left( \bm{x}^{\prime \prime}_r \right) \right \rangle + U \left( x^{\prime \prime}_r \right) \frac{\partial}{\partial x^\prime_r} \left \langle H^{(1)} \left( \bm{x}^\prime_r \right) Y \left( \bm{x}^{\prime \prime}_r \right) \right \rangle \right ].
\label{cvu}
\end{align}

Thus, central for the present study is the fluctuation~$H^{(1)}$ that is derived as:
\begin{equation} \label{Hn1}
H^{(1)} \left( \bm x \right) = Q \int \dd \bar{\bm x} \, G^\infty_3 \left( \bm x - \bar{\bm x} \right) \frac{\partial Y \left( \bar{\bm x} \right)}{\partial \bar{x}_m}  \frac{\partial G^\infty_2 \left( \bar{\bm x}_r \right)}{\partial \bar{x}_m} \qquad \quad \left( m = 1,2 \right)
\end{equation}
\citep{Fiori1998}. It is convenient to write the head's fluctuation~\eqref{Hn1} as~$H^{(1)} \left( \bm{x} \right) = Q /\left( 2 \pi \right)^{3/2} \int \dd \bm{k} \tilde{Y}(\bm{k}) \exp \left( - \jmath \bm{k} \cdot \bm x \right) \mathcal H \left( \bm k \right)$ with
\begin{equation}\label{Hnn}
\mathcal H \left( \bm k \right) = - \jmath k_m \int \dd \bar{\bm x} \exp \left( - \jmath \bm{k} \cdot \bar{\bm x} \right) G^\infty_3 \left( \bar x \right) \frac{\partial}{\partial \bar{x}_m}  G^\infty_2 \left( | \bm{x}_r - \bm{\bar x}_r | \right),
\end{equation}
where the fluctuation~$Y$ has been written by means of its spectral (\textit{Fourier transform}) representation~$\tilde Y$, \ie
\begin{equation} \label{FT}
Y (\bm{x}) = \int \frac{\dd \bm{k}}{\left( 2 \pi \right)^{3/2}}  \, \tilde{Y} \left( \bm{k} \right) \exp \left( - \jmath \bm{x} \cdot \bm{k} \right).
\end{equation}
It is therefore clear from~\eqref{Hnn} that, similarly to the other, above cited applications, even the problem of modelling of scattering through randomly heterogeneous porous media~\textit{de facto} calls for the computation of~\eqref{model}, with~$a_m$ replaced by~$-\jmath k_m$. \\
\indent The remainder of the paper is organized as follows: we compute explicitly the integral~\eqref{Hnn}. Then, we discuss the structure and the behavior of the flow variables related to it, before moving to the modelling of scattering through heterogeneous porous formations. Finally, we end up with concluding remarks.

\section*{Analytical computation of~$\mathcal H \equiv \mathcal H \left( \bm a \right)$}
A direct computation of~\eqref{Hnn} does not seem achievable, unless one deals with particular structures of heterogeneity~\citep{severino2011}. For this reason we follow in the sequel a different avenue. More precisely, we start from the unsteady state version of the same problem, \ie
\begin{equation}\label{nde}
\exp \left( - Y \right) \frac{\partial}{\partial t} G - \nabla^2
 G - \nabla Y \cdot \nabla G = \delta(\bm{x}_r) \delta \left( t \right), \qquad \quad G \left( \bm x, 0\right) = 0,
\end{equation}
and compute the integral~\eqref{Hn1} as~$\displaystyle \lim_{t \to \infty} \int^t_0 \dd \tau \, G^{(1)} \left( \bm x, \tau \right)$, by virtue of the~\textit{superposition principle}, being~$G^{(1)} \equiv G^{(1)} \left( \bm x, t \right)$ the first order approximation of~\eqref{nde}. In particular, for a homogeneous medium  (\ie~$Y \equiv 0$) one recovers  from~\eqref{nde} the equation of the~$d$-dimensional Green function, \ie~$G_d \left(\bm x, t \right) = \left( 4 \pi t \right)^{-d/2} \exp \left[ - |\bm x|^2 / \left( 4t \right) \right]$. In order to compute~$G^{(1)}$, we procede like before. Thus, we expand~$G$ in the asymptotic series~$G = G^{(0)} + G^{(1)} + \dots$ with~$G^{(n)} = \mathcal{O} \left( Y^n \right)$. Then, substitution into~\eqref{nde} and retaining the first order term provide the equation for the fluctuation~$G^{(1)}$, \ie
\begin{equation}\label{nde1}
\frac{\partial}{\partial t}G^{(1)} - \nabla^2 G^{(1)} = Y \frac{\partial}{\partial t} G^{(0)} + \nabla Y \cdot \nabla G^{(0)}, \qquad \quad G^{(0)} \equiv G_3.
\end{equation}
To solve eq.~\eqref{nde1}, we apply Laplace transform over the time, and Fourier transform~\eqref{FT} over the space. The final result, after employing integration by parts, reads as:

\begin{align} 
& G^{(1)}(\bm{x},t) = - \int \frac{\dd \bm{k} \, \tilde{Y} \left( \bm{k} \right)}{\left( 2 \pi \right)^{3/2}} \int^t_0 \dd \tau \int \dd \bar{\bm{x}} \exp \left( - \jmath \, \bm{k} \cdot \bar{\bm{x}} \right) \bigg [ \delta \left( \bar{\bm{x}}_r \right) \delta \left( \tau \right) G_3 \left( |\bm x - \bar{\bm x}| , t - \tau \right) -  \notag \\
& \frac{\partial}{\partial \bar{x}_m} G_3 \left( |\bm x - \bar{\bm x}| , t - \tau \right) \frac{\partial}{\partial \bar{x}_m} G_2(\bar{x}_r, \tau) \bigg ] = \jmath k_m \int \frac{\dd \bm{k} \, \tilde{Y} \left( \bm{k} \right)}{\left( 2 \pi \right)^{3/2}} \int^t_0 \dd \tau \int \dd \bar{\bm{x}} \exp \left( - \jmath \, \bm{k} \cdot \bar{\bm{x}} \right) \times \notag \\
& G_3 \left( |\bm x - \bar{\bm x}| , t - \tau \right) \frac{\partial}{\partial \bar{x}_m} G_2(\bar{x}_r, \tau) \qquad \qquad \left ( m=1,2 \right).
\label{firstnew}
\end{align}
We now compute the inner (spatial) quadratures appearing into the last of~\eqref{firstnew}, \ie

\begin{align}
& \jmath k_m \int^t_0 \dd \tau \int \dd \bar{\bm{x}} \exp \left( - \jmath \, \bm{k} \cdot \bar{\bm{x}} \right) G_3 \left( |\bm x - \bar{\bm x}| , t - \tau \right) \frac{\partial}{\partial \bar{x}_m} G_2(\bar{x}_r, \tau) = - \frac{\jmath}{2} \exp \left( - \jmath k_3 x_3 \right) \times \notag \\
& \int^t_0 \frac{\dd \tau}{\tau} \exp \left[ - k^2_3 \left( t - \tau \right) \right] \int \dd \bar{\bm{x}}_r \exp \left( - \jmath \, \bm{k}_r \cdot \bar{\bm{x}}_r \right) G_2 \left( |\bm x_r - \bar{\bm x}_r | , t - \tau \right) \bm{k}_r \cdot \bar{\bm{x}}_r \, G_2(\bar{x}_r, \tau) = \notag \\
& \left ( 8 \pi \right)^{-1} \exp \left( - \jmath k_3 x_3 \right) \lim_{\alpha \to \jmath} \int^t_0 \frac{\dd \tau}{\tau^2} \exp \left[ - k^2_3 \left( t - \tau \right) \right] G_2 \left( x_r, t - \tau \right) \alpha \frac{\partial}{\partial \alpha} \, \mathcal{I} \left( \alpha \right),
\label{A1} 
\end{align}
where we have set
\begin{equation} \label{Ia}
\mathcal{I} \left( \alpha \right) = \int \dd \bar{\bm{x}}_r \exp \left( - a \bar{x}^2_r \right) \exp \left( \bm{\omega}_\alpha \cdot \bar{\bm{x}}_r \right),
\end{equation}
being~$\bm{\omega}_\alpha \equiv b \bm x_r -\alpha \bm k_r$,~$a \equiv \dfrac{t}{4 \left( t - \tau \right) \tau}$ and~$b \equiv \dfrac{1}{2 \left( t - \tau \right)}$.
The evaluation of~$\mathcal{I} \left( \alpha \right)$ is straightforward. By skipping the algebraic details, it yields~$\mathcal{I} \left( \alpha \right) = \left( \pi / a \right) \exp \left[ \bm{\omega}_\alpha \cdot \bm{\omega}_\alpha /\left( 4 a \right) \right]$. As a consequence, eq.~\eqref{firstnew} writes as:
\begin{equation}
G^{(1)}(\bm{x},t) = - \frac{\jmath}{2t} G_2 \left( x_r, t \right) \int \frac{\dd \bm{k} \tilde{Y} \left( \bm{k} \right)}{\left( 2 \pi \right)^{3/2}} \exp \left( - \jmath k_3 x_3 \right) \int^t_0 \dd \tau \Gamma \left( \tau \right) \exp \left [ \jmath \frac{t - \tau}{t} \left( \jmath \tau \bm k_r - \bm x_r \right) \cdot \bm k_r \right ],
\label{G1}
\end{equation}
with~$\Gamma \left( t \right) \equiv \left ( \bm x_r \cdot \bm k_r - 2 \jmath t k^2_r \right) \exp \left (-k^2_3 t \right)$. We are now in position to calculate the fluctuation~$h^{(1)}(\bm{x},t) = \displaystyle \int^t_0 \dd \tau \, G^{(1)}(\bm{x},\tau)$ that, after changing the order of integration and performing one quadrature, becomes:

\begin{align}
& h^{(1)}(\bm{x},t) = - \frac{\jmath}{8 \pi} \int \frac{\dd \bm{k} \, \tilde{Y} \left( \bm{k} \right)}{\left( 2 \pi \right)^{3/2}} \exp \left( - \jmath k_3 x_3 \right) \exp \left( -\jmath \bm x_r \cdot \bm k_r  \right) \int^t_0 \dd \tau^\prime \, \Gamma \left( \tau^\prime \right) \exp \left( - k^2_r \, \tau^\prime \right) \times \notag \\
& \int^t_{\tau^\prime} \frac{\dd \tau^{\prime \prime}}{\tau^{\prime \prime \, 2}} \exp \left( - \frac{\omega_{\tau^\prime}}{4 \tau^{\prime \prime}} \right) = - \frac{\jmath}{2 \pi} \, \int \frac{\dd \bm{k} \, \tilde{Y} \left( \bm{k} \right)}{\left( 2 \pi \right)^{3/2}} \exp \left( - \jmath k_3 x_3 \right) \exp \left( - \jmath \bm x_r \cdot \bm k_r  \right) \int^t_0 \dd \tau \exp \left( -k^2 \tau \right) \notag \\
& \times \beta (\tau) \left[ \exp \left( - \frac{\omega_\tau}{4 u} \right) \right]^{u = t}_{u = \tau}, \quad \omega_t \equiv x^2_r + 4 \jmath t \left( \jmath t \bm k_r - \bm x_r \right) \cdot \bm k_r, \, \, \, \, \beta \left( t \right) = \frac{\bm k_r}{\omega_t} \cdot \left( \bm x_r - 2 \jmath t \bm k_r \right). 
\label{ff}
\end{align}

As above anticipated, we now focus on the large time behavior of~\eqref{ff}. Toward this aim, we preliminarily note that, for~$t \gg 1$, the dominant contribution in the integrand of~\eqref{ff} (that is achieved upon asymptotic expansion, and by retaining the leading order term) is such that:
\begin{equation} \label{asympt}
\beta \left( \tau \right) \simeq \frac{\jmath}{2 \tau}, \qquad \qquad \exp \left ( - \frac{\omega_\tau}{4t} \right ) \simeq \exp \left ( - \frac{x^2_r}{4 \tau} \right).
\end{equation}
Hence, by replacing the functions~$\beta (\tau)$ and~$\exp \left[ - \omega_\tau / (4t) \right]$ with the approximations~\eqref{asympt} leads to:

\begin{align}
& h^{(1)}(\bm{x},t) = - \frac{\jmath}{2 \pi} \, \int \frac{\dd \bm{k} \, \tilde{Y} \left( \bm{k} \right)}{\left( 2 \pi \right)^{3/2}} \exp \left( - \jmath k_3 x_3 \right) \Bigg [ \exp \left( -\jmath \bm x_r \cdot \bm k_r  \right) \int^t_0 \dd \tau \, \beta \left( \tau \right) \, \exp \left( - k^2 \tau - \frac{\omega_\tau}{4 t} \right) \notag \\
& - \int^t_0 \dd \tau \, \beta \left( \tau \right) \, \exp \left( - k^2_3 \tau - \frac{x^2_r}{4 \tau} \right) \Bigg ] \simeq \frac{1}{4 \pi} \int \frac{\dd \bm{k} \, \tilde{Y} \left( \bm{k} \right)}{\left( 2 \pi \right)^{3/2}} \exp \left( - \jmath k_3 x_3 \right) \Bigg [ \exp \left( -\jmath \bm x_r \cdot \bm k_r  \right) \times \notag \\
& \int^t_0 \frac{\dd \tau}{\tau} \exp \left( - k^2 \tau - \frac{x^2_r}{4 \tau} \right)  - \int^t_0 \frac{\dd \tau}{\tau} \exp \left( - k_3^2 \tau - \frac{x^2_r}{4 \tau} \right) \Bigg] + \mathcal{O} \left( t^{-1} \right).
\label{ffn}
\end{align}

Finally, by taking the limit~$t \to \infty$ in the last of~\eqref{ffn} gives:
\begin{equation} \label{ff9}
H^{(1)}(\bm{x}) = \lim_{t \to \infty} h^{(1)}(\bm{x},t) = \int \frac{\dd \bm{k} \, \tilde{Y} \left( \bm{k} \right)}{\left( 2 \pi \right)^{5/2}} \exp \left( - \jmath k_3 x_3 \right) \left [ \exp \left( -\jmath \bm x_r \cdot \bm k_r  \right) \mathrm{K}_0 (x_r k) - \mathrm{K}_0 (x_r |k_3|) \right],
\end{equation}
where~$\mathrm{K}_n$ is the $n$-order modified Bessel function of first kind. The comparison of~\eqref{ff9} with~\eqref{Hnn} suggests that the wide concern integral~\eqref{model} is equal to:
\begin{equation} \label{H}
\mathcal H \left( \bm a \right) = \left( 2 \pi \right)^{-1} \left[ \mathrm{K}_0 (|\bm a| x_r) - \exp \left( - \bm{a}_r \cdot \bm{x}_r \right) \mathrm{K}_0 (|a_3| x_r) \right].
\end{equation}
\begin{figure}
\vspace{-5 pt}
\includegraphics[width=1.0\textwidth]{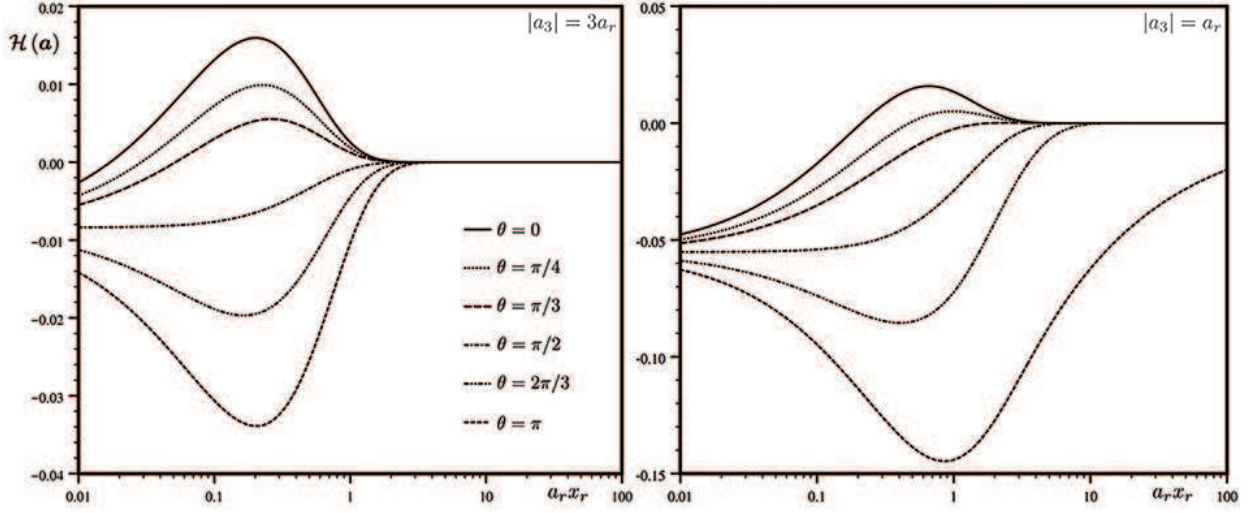}
\caption{Dependence of the function~$\mathcal H \equiv \mathcal H \left( \bm a \right)$ upon the nondimensional distance~$a_r x_r$ and polar angles~$\theta = \arccos \left[ \bm a_r \cdot \bm x_r /(a_r x_r) \right]$. Other values:~$|a_3| = 3a_r$ and~$|a_3| = a_r$.} \label{Figure1}
\end{figure}
For illustration purposes, the function~\eqref{H} is depicted in FIG.~\ref{Figure1} versus the dimensionless variable~$a_r x_r$ (with $\bm a_r \in \mathbb{R}^2$), a few values of the polar angle~$\theta = \arccos \left[ \bm a_r \cdot \bm x_r /(a_r x_r) \right]$ and two values of~$|a_3|$. The quantity~$\displaystyle \lim_{x_r \to 0^+} \mathcal H = \left( 2 \pi \right)^{-1} \ln \left ( | a_3 | / | \bm a | \right)$ is instrumental in the engineering applications, in order to let the head's fluctuation meet a Dirichlet boundary condition at the source (\textit{regularization}). At the other extreme of large distances, the function~\eqref{H} vanishes with exponential decay. In what follows, we proceed with analyzing second-order moments of the flow variables that, under the above stated conditions, result of the same order of magnitude of the~$Y$-$\,$variance~$\sigma^2_Y = \langle Y^2 \rangle$.

\section*{Discussion}
We wish to derive and discuss some statistical parameters that quantify the uncertainty in the spatial distribution of the specific energy~$H$ and the velocity~$\bm V$. Starting with the cross-covariance~$C_{YH} \left( \bm x, \bm y \right) \equiv \left \langle Y \left( \bm x \right) H^{(1)} \left( \bm y \right) \right \rangle$, it results from~\eqref{ff9} as:
\begin{equation} \label{cYH}
\frac{C_{YH} (\bm x, \bm y)}{Q\sigma^2_Y} = \int \frac{\dd \bm k \, \tilde \rho_Y \left( \bm k \right)}{\left( 2 \pi \right)^{5/2}} \exp \left( \jmath \xi_3 k_3 \right) \left [ \exp \left( \jmath \bm \xi_r \cdot \bm k_r  \right) \mathrm{K}_0 (y_r k) - \exp \left( \jmath \bm x_r \cdot \bm k_r  \right)\mathrm{K}_0 (y_r |k_3|) \right]
\end{equation}
$\left( \bm \xi \equiv \bm x - \bm y \right)$, where we have made use of the stationarity of~$Y$, \ie
\begin{equation}
\left \langle \tilde{Y} \left( \bm{k}_1 \right) \tilde{Y} \left( \bm{k}_2 \right) \right \rangle = \left( 2 \pi \right)^{3/2} \sigma^2_Y \, \delta \left( \bm{k}_1 + \bm{k}_2 \right) \tilde \rho_Y \left( \bm k_2 \right).
\end{equation}
Likewise, the head covariance~$C_H (\bm x, \bm y) \equiv \langle H^{(1)}(\bm x) H^{(1)}(\bm y) \rangle$ is obtained by multiplying~\eqref{ff9} applied at
two points~$\bm x \neq \bm y$, and subsequently taking the ensemble average. The final result is:

\begin{align}
\frac{C_{H} (\bm x , \bm y)}{\left( Q \sigma_Y \right)^2} = & \int \frac{\dd \bm k \, \tilde \rho_Y \left( \bm k \right)}{\left( 2 \pi \right)^{7/2}} \exp \left( \jmath \xi_3 k_3 \right) \big[ \exp \left( - \jmath \bm \xi_r \cdot \bm k_r \right) \mathrm{K}_0 (x_r k) \mathrm{K}_0 (y_r k) + \mathrm{K}_0 (x_r |k_3|)\mathrm{K}_0 (y_r |k_3|) \notag \\
& - \exp \left( - \jmath \bm x_r \cdot \bm k_r \right) \mathrm{K}_0 (x_r k) \mathrm{K}_0 (y_r |k_3|) - \exp \left( \jmath \bm y_r \cdot \bm k_r \right) \mathrm{K}_0 (y_r k) \mathrm{K}_0 (x_r |k_3|) \big].
\label{CH}
\end{align}
It is seen that the covariances~\eqref{cYH} and~\eqref{CH} are stationary along the vertical coordinate (\ie they depend only upon the lag $\xi_3 = x_3 - y_3$), since the mean value~$H^{(0)} \left( x_r \right) \equiv Q \, G_2^\infty \left( x_r \right)$ does not depend upon the elevation. Moreover, based on the existing data-sets~\citep[an exhaustive overview can be found in][]{rubin2003}, we regard the autocorrelation of~$Y$ as axial symmetric, and therefore the spectrum~$\tilde \rho_Y \left( \bm k \right) \equiv \tilde \rho_Y (k_r, k_3)$ is an even function of~$k_r$ and~$k_3$. Hence, by adopting cylindrical coordinates in wave-number space, \ie~$\bm k \equiv \left( k_r \cos \theta, k_r \sin \theta, k_3 \right)$, and carrying out the quadrature over the polar angle lead to:
\begin{equation} \label{CYHN}
\frac{C_{YH} (\bm x, \bm y)}{Q\sigma^2_Y} = 2 \int^\infty_0 \int^\infty_0 \frac{\dd k_r \dd k_3}{\left( 2 \pi \right)^{3/2}} \, k_r \, \tilde \rho_Y \left( k_r, k_3 \right) \cos \left( \xi_3 k_3 \right) \left [ J_0 (\xi_r k_r) \mathrm{K}_0 (y_r k) - J_0 (x_r k_r) \mathrm{K}_0 (y_r k_3) \right],
\end{equation}
\begin{align}
& \frac{C_{H} (\bm x , \bm y)}{\left( Q \sigma_Y \right)^2} = 2 \int^\infty_0 \int^\infty_0 \frac{\dd k_r \dd k_3}{\left( 2 \pi \right)^{5/2}} \, k_r \, \tilde \rho_Y \left( k_r, k_3 \right) \cos \left( \xi_3 k_3 \right) \big[ J_0 \left( \xi_r  k_r \right) \mathrm{K}_0 (x_r k) \mathrm{K}_0 (y_r k) \notag \\
& + \mathrm{K}_0 (x_r k_3)\mathrm{K}_0 (y_r k_3) - J_0 \left( x_r k_r \right) \mathrm{K}_0 (x_r k) \mathrm{K}_0 (y_r k_3) - J_0 \left( y_r k_r \right) \mathrm{K}_0 (y_r k) \mathrm{K}_0 (x_r k_3) \big]
\label{CHN}
\end{align}
($J_n$ is the $n$-order Bessel function of the first kind). Two parameters are of particular interest, namely the cross,~$\sigma_{YH} \left( x_r \right) \equiv C_{YH} (\bm x, \bm x)$, and the head,~$\sigma^2_H \left( x_r \right) \equiv C_H (\bm x, \bm x)$, variances which are derived from~\eqref{CYHN}--\eqref{CHN} as follows:
\begin{equation} \label{sYH}
\frac{\sigma_{YH} \left( x_r \right)}{Q\sigma^2_Y} = 2 \int^\infty_0 \int^\infty_0 \frac{\dd k_r \, \dd k_3}{\left( 2 \pi \right)^{3/2}} \, k_r \, \tilde \rho_Y \left( k_r, k_3 \right)  \left [ \mathrm{K}_0 (x_r k) - J_0 (x_r k_r) \mathrm{K}_0 (x_r k_3) \right],
\end{equation}
\begin{equation} \label{sH}
\frac{\sigma^2_H (x_r)}{\left( Q \sigma_Y \right)^2} = 2 \int^\infty_0 \int^\infty_0 \frac{\dd k_r \dd k_3}{\left( 2 \pi \right)^{5/2}} k_r \tilde \rho_Y \left( k_r, k_3 \right) \left[ \mathrm{K}^2_0 (x_r k) + \mathrm{K}^2_0 (x_r k_3) - 2 J_0 \left( x_r k_r \right) \mathrm{K}_0 (x_r k) \mathrm{K}_0 (x_r k_3) \right].
\end{equation}
\begin{figure}
\includegraphics[width=1\textwidth]{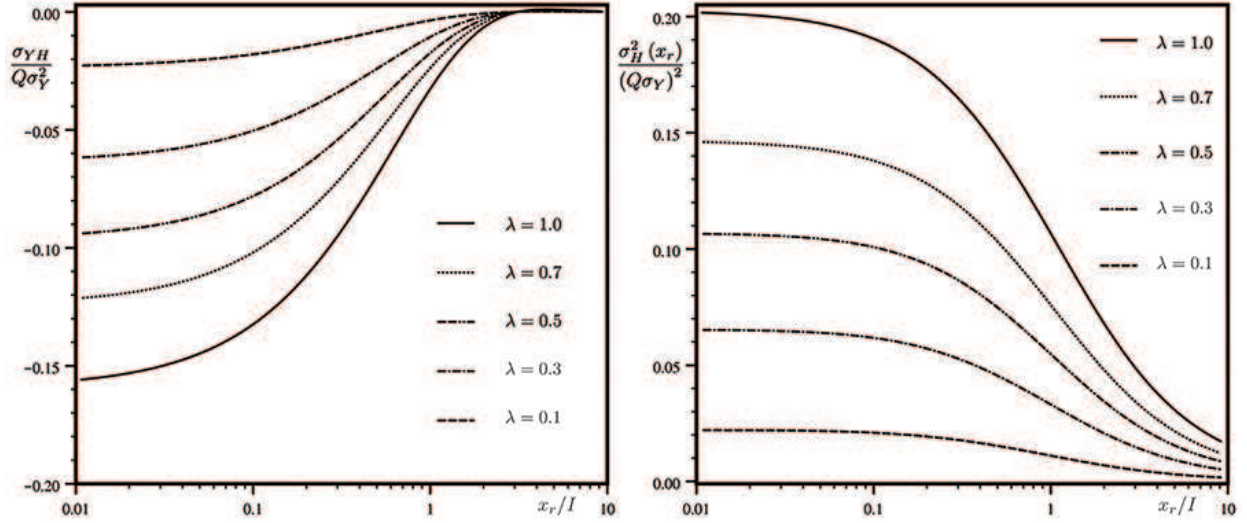}
\caption{Dependence of the scaled cross-variance~$\sigma_{YH} / (Q \sigma^2_Y)$ and variance~$\sigma^2_H / (Q \sigma_Y)^2$ upon the dimensionless distance~$x_r / I$ from the source, and several values of the anisotropy ratio~$\lambda$ (exponential spectrum of~$\rho_Y$).}
\label{Figure2}
\end{figure}
\indent To explore the physical insights of eqs~\eqref{sYH}--\eqref{sH}, we adopt an exponential model for the spectrum, \ie~$\tilde \rho_Y (k_r, k_3) \equiv \left( 8 / \pi \right)^{1/2} \lambda \left( 1 + k^2_r + \lambda^2 k^2_3 \right)^{-2}$, where the anisotropy ratio~$\lambda \in ]0,1]$ is defined as the ratio between the vertical, \ie~$I_v$, and horizontal, \ie~$I$, integral scales of~$Y$. In addition, the wave numbers~$\left (\bm k_r, k_3 \right)$ have been made dimensionless by replacing $k_i \to I k_i$ (with~$\ell_c \equiv I$). In FIG.~\ref{Figure2} the cross-variance~\eqref{sYH} is depicted as function of the scaled variable~$x_r / I$ and a few values of~$\lambda$. It is a monotonic increasing function of~$x_r$ that starts from the value at the source, \ie
\begin{equation}
\sigma_{YH} \left( 0 \right) = \frac{2 Q\sigma^2_Y}{\left( 2 \pi \right)^{3/2}} \int^\infty_0 \int^\infty_0 \dd k_r \dd k_3 \, k_r \, \tilde \rho_Y \left( k_r, k_3 \right) \ln \frac{k_3}{k} = - Q \sigma^2_Y \frac{\lambda}{2 \pi } \frac{\arcsin \sqrt{1-\lambda^2}}{\sqrt{1-\lambda^2}},
\label{near}
\end{equation}
and it vanishes after four horizontal integral scales. In particular, the near field~\eqref{near} is valid also for Gaussian spectrum:~$\tilde \rho_Y (k_r, k_3) \equiv \left( 2 / \pi \right)^{3/2} \lambda \exp \left( - k^2_r/ \pi - \lambda^2 k^2_3 / \pi \right)$. \\
\indent In order to explain the behavior of the cross-variance~$\sigma_{YH}$, 
\begin{figure}
\vspace{-0.25 cm}
\centering
\includegraphics[width=0.85\textwidth]{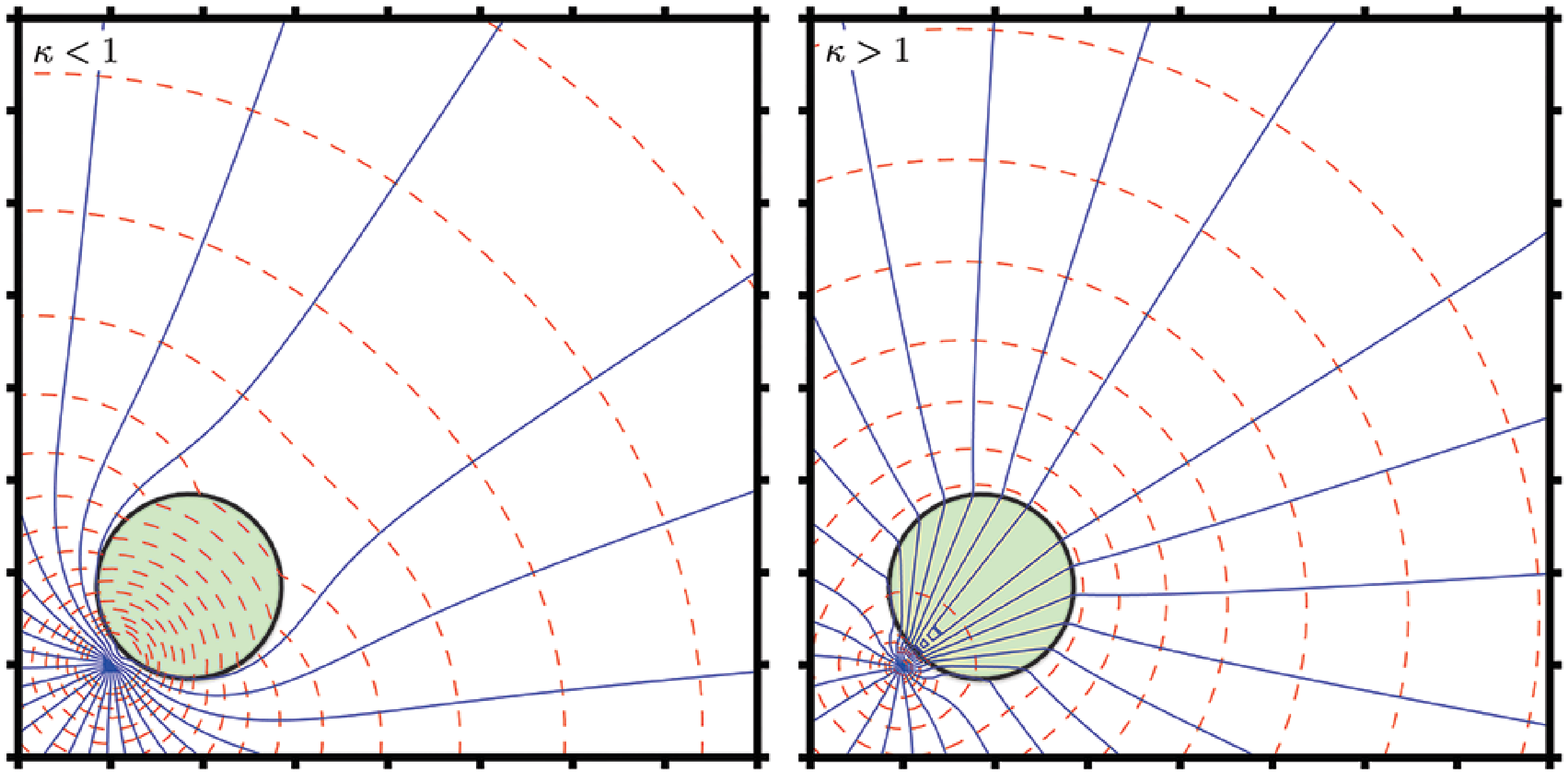}
\vskip 3 mm
\includegraphics[width=0.85\textwidth]{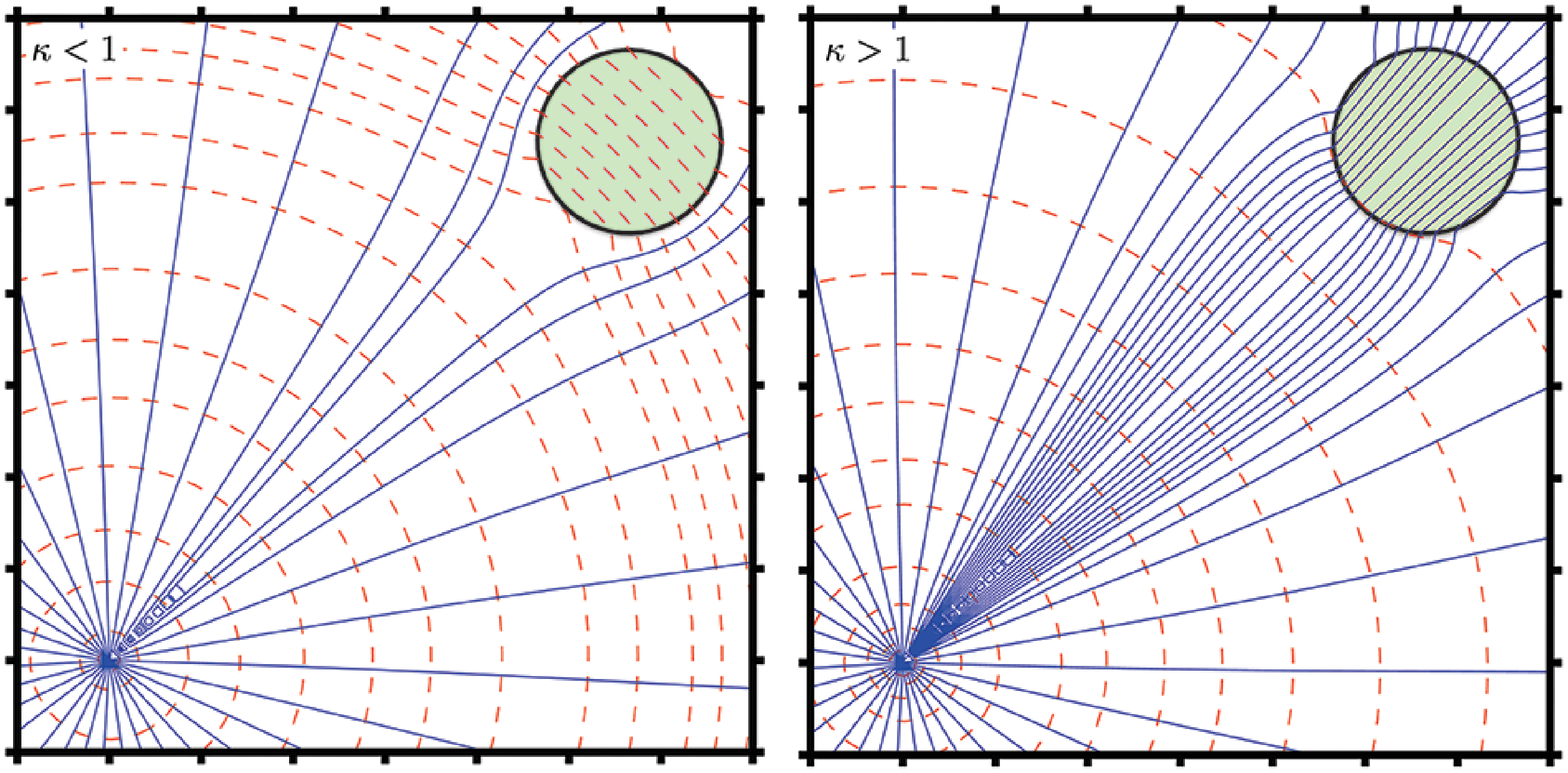}
\caption{Contour-plot of the head (red dashed lines) and stream function (blue continuous lines) as affected by a circular (green) inclusion of conductivity $K$ implanted into a matrix of effective conductivity $K_{\rm{eff}}$. On the top, pictures refer to an inclusion close to the source with contrast ratio~$\kappa = K / K_{\rm{eff}}$ smaller and larger than one. Below, pictures pertain to the analogous situation, but for an inclusion lying away from the source.}
\label{Figure3}
\end{figure}
we can focus on the flow's pattern as deformed by a single inclusion of conductivity $K$~\citep{severino2019effective} embedded into a matrix of effective conductivity $K_{\rm eff}$ (being~$\sigma_{YH}$ evaluated as average of the product between the fluctuations $H^{(1)}$ and~$Y$ over many of such realizations). Thus, in the~FIG.~\ref{Figure3} we have depicted a circular (green) inclusion near and far from the source for two largely different values of the~\textit{contrast ratio} $\kappa \equiv K / K_{\rm eff}$. In particular, due to the mass conservation, the streamlines circumvent the inclusion for $\kappa < 1$ and they are attracted by it for $\kappa > 1$. As a consequence, in the near and far field the head surrounding the inclusion results higher/lower than the mean head (corresponding to~$\kappa = 1$) for~$\kappa <1$ and~$\kappa > 1$, respectively. Thus, for~$\kappa < 1$ (calling for~$Y < 0$) the fluctuation~$H^{(1)}$ is larger than the mean, and~\textit{viceversa}. Hence, the product~$Y \left( x_r \right) H^{(1)} \left( x_r \right)$ (and concurrently the ensemble average~$\sigma_{YH}$) results lesser than zero, in any case. The limit~$\displaystyle \lim_{x_r \to \infty} \sigma_{YH} \left( x_r \right) = 0$ is explained by recalling that the head's fluctuation tends to zero away from the source (see~\eqref{ff9}). Finally, the reduction of~$\sigma_{YH}$ (for given~$x_r$) with increasing~$\lambda$ has a straightforward kinematical reasoning: an anisotropic medium can be sought as made up by inclusions elongated in the horizontal direction (resembling the medium's structure~$\lambda \equiv I_v / I < 1$). Thus, for a fluid particle it is easier to circumvent a low conducting inclusion by moving vertically rather than laterally. This causes a deviation from the mean lesser than that which one would observe within a medium of isotropic ($\lambda = 1$) heterogeneity's structure. \\
\indent By the same token, one can analyze the scaled variance~$\sigma^2_H /\left( Q \sigma_Y \right)^2$. Thus, at large~$x_r$ the head is quite small, since the flow there behaves as a homogeneous one~\citep{abramovich1995}, which decays like~$x^{-1}_r$. To the contrary, in the region close to the source the mean head~$H^{(0)}$ is highly uncertain, since most of the head buildup takes place within a tiny annulus surrounding the source~\citep{severino2019uncertainty}. The dependence of~$\sigma^2_H$ upon the anisotropy ratio~$\lambda$ (at any given distance) is explained by the same argument as before. \\
\indent The variance~$\sigma^2_u \left( x_r \right) \equiv u_{rr} \left( x_r, x_r \right)$ of the velocity is obtained from~\eqref{cvu} as:
\begin{equation}
\sigma^2_u \left( x_r \right) = \sigma^2_Y \, U^2 \left( x_r \right) + 2 \, \frac{K_G}{n} \, U \left( x_r \right) \sigma_{YE_r} \left( x_r \right) + \left ( \frac{K_G}{n} \right)^2 \sigma^2_{E_r} \left( x_r \right), \quad E_r \equiv \frac{\partial}{\partial x_r} H^{(1)} \left( \bm x \right),
\end{equation}
where we have set~$\sigma_{YE_r} \equiv \left \langle Y  E_r \right \rangle$ and~$\sigma^2_{E_r} \equiv \left \langle E^2_r \right \rangle$. By differentiation of~\eqref{ff9}, the latter are given by:
\begin{equation}
\sigma_{YE_r} \left( x_r \right) = \frac{2 Q \sigma^2_Y}{\left( 2 \pi \right)^{3/2}} \int^\infty_0 \int^\infty_0 \dd k_r \, \dd k_3 \, k_r \, \tilde \rho_Y \left( k_r, k_3 \right)  \left [ k_3 J_0 (x_r k_r) \mathrm{K}_1 (x_r k_3) - k \mathrm{K}_1 (x_r k) \right],
\end{equation}
\begin{align}
& \sigma^2_{E_r} \left( x_r \right) = \frac{\left (Q \sigma_Y \right)^2}{\left( 2 \pi \right)^{5/2}} \int^\infty_0 \int^\infty_0 \dd k_r \, \dd k_3 \, k_r \, \tilde \rho_Y \left( k_r, k_3 \right) \big \{ 2 \left [ k \mathrm{K}_1 \left( x_r k \right) \right]^2 + 2 \left [ k_3 \mathrm{K}_1 \left( x_r k_3 \right) \right]^2 - \notag \\
& \left [ k_r \mathrm{K}_0 \left( x_r k \right) \right]^2 - 2 k_3 \mathrm{K}_1 \left( x_r k_3 \right) \left[ k_r J_1 \left( x_r k_r \right) + 2 k J_0 \left( x_r k_r \right) \mathrm{K}_1 \left( x_r k \right) + k_r J_1 \left( x_r k_r \right) \mathrm{K}_0 \left( x_r k \right) \right] \big \}.
\end{align}

\begin{figure} [h]
\includegraphics[width=1.0 \textwidth]{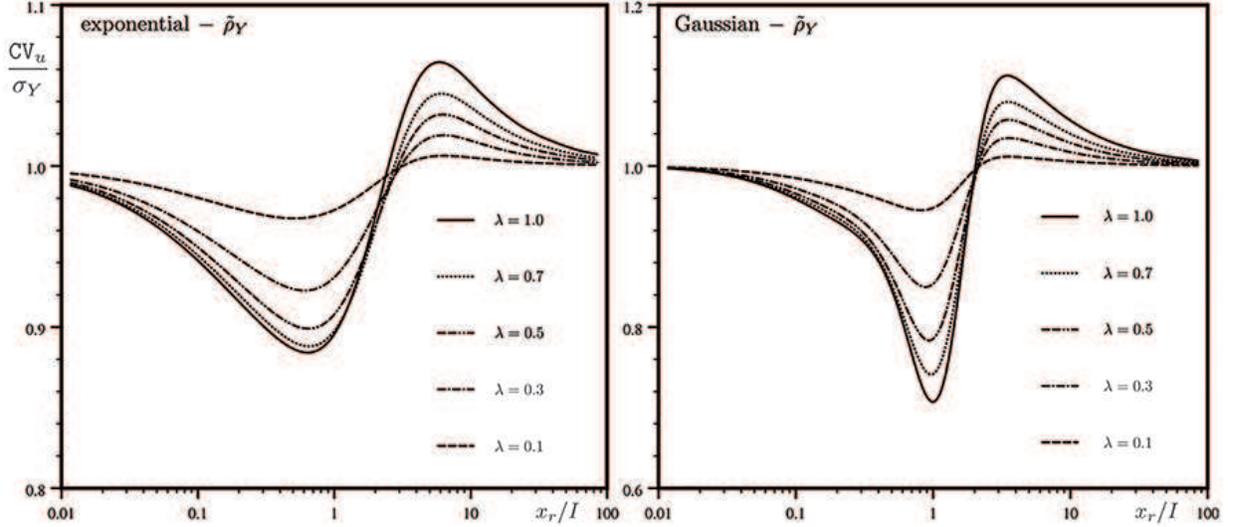}
  \caption{Scaled coefficient of variation~$\texttt{CV}_u/ \sigma_Y$ versus the normalized distance~$x_r / I$ from the source, and a few values of the anisotropy ratio~$\lambda$ (exponential and Gaussian spectrum).}
\label{Figure4}
\end{figure}

The scaled coefficient of variation~$\texttt{CV}_u /\sigma_Y = \sigma_u / \left ( U \sigma_Y \right)$ is depicted (for both exponential and Gaussian~$\tilde{\rho}_Y$) in the FIG.~\ref{Figure4}.
It is seen that in the near (\ie~$x_r \ll I$) and far (\ie~$x_r \gg I$) field, one has~$\sigma_u \sim \sigma_Y U$. Indeed, close to the source the flow can be homogenized by the harmonic (constant) conductivity~\citep{indelman1996averaging}, whereas far from the source it behaves like a mean uniform one of effective conductivity. As a consequence, in these two regimes the uncertainty in the velocity field resembles precisely the reduction of the mean velocity~$U$ with the distance. In the intermediate regime, for~$x_r < I$ the cross-variance (that is negative) is mostly influential, and concurrently~$\texttt{CV}_u$ reduces, whereas for~$x_r > I$ it rapidly exhausts, with a still impact of the head-gradient's variance~$\sigma^2_{E_r}$. This justifies the sudden rise of~$\texttt{CV}_u$. As it will be clearer later on, these findings are of paramount importance when analyzing the evolution of scattering. To conclude this section, we note that the Gaussian shape of~$\rho_Y$ produces a more persistent signal in the coefficient of variation of the velocity~\citep[in agreement with][]{severino2020uncertainty}.

\section*{Scattering analysis}
We are now in position to analyze scattering of a passive scalar as determined by the above discussed source-type flow. This goal is achieved by means of the second-order radial moment~\eqref{Xrrr} which, for convenience of discussion, is re-written on the base of~\eqref{cvu} as:
\begin{equation} \label{dec}
X_{rr} \left( R \right) =  \mathcal{X}_\infty \left( R \right) + \mathcal{X}_\star \left( R \right),
\end{equation}
being
\begin{equation}
\mathcal{X}_\infty \left( R \right) =  \sigma^2_Y \, U^2 \left( R \right) \int^R_0 \int^R_0 \dd x^\prime_r \, \dd x^{\prime \prime}_r \, \frac{\rho_Y \left( x^\prime_r - x^{\prime \prime}_r \right)}{U \left( x^\prime_r \right) U \left( x^{\prime \prime}_r \right)} = \frac{\sigma^2_Y}{3} \, R \int^R_0 \dd u \left( 2 - 3 \frac{u}{R} + \frac{u^3}{R^3} \right) \rho_Y \left( u \right),
\label{Xinf}
\end{equation}
whereas

\begin{align}
& \mathcal{X}_\star \left( R \right) = \, U^2 \left( R \right) \int^R_0 \int^R_0  \frac{\dd x^\prime_r \, \dd x^{\prime \prime}_r}{U^2 \left( x^\prime_r \right) U^2 \left( x^{\prime \prime}_r \right)} \Bigg[ \left( \frac{K_G}{n} \right)^2 \frac{\partial^2}{\partial x^\prime_r \partial x^{\prime \prime}_r} \left \langle H^{(1)} \left( x^\prime_r \right) H^{(1)} \left( x^{\prime \prime}_r \right) \right \rangle - \notag \\
& \frac{K_G}{n} \, U \left( x^\prime_r \right) \frac{\partial}{\partial x^{\prime \prime}_r} \left \langle Y \left( x^\prime_r \right) H^{(1)} \left( x^{\prime \prime}_r \right)  \right \rangle - \frac{K_G}{n} \, U \left( x^{\prime \prime}_r \right) \frac{\partial}{\partial x^\prime_r} \left \langle H^{(1)} \left( x^\prime_r \right) Y \left( x^{\prime \prime}_r \right) \right \rangle  \Bigg ] = \frac{K_G}{n} \, \, U^2 \left( R \right) \times \notag \\
& \int^R_0 \int^R_0  \frac{\dd x^\prime_r \, \dd x^{\prime \prime}_r}{U^2 \left( x^\prime_r \right) U^2 \left( x^{\prime \prime}_r \right)} \left[ \left( \frac{K_G}{n} \right) \frac{\partial^2 \, C_H \left( x^\prime_r, x^{\prime \prime}_r \right)}{\partial x^\prime_r \partial x^{\prime \prime}_r} - 2 \, U \left( x^\prime_r \right) \frac{\partial \, C_{YH} \left( x^\prime_r, x^{\prime \prime}_r \right) }{\partial x^{\prime \prime}_r} \right ].
\label{Xstar}
\end{align}
In particular, the last of~\eqref{Xstar} has been achieved by noting that~$\left(x^\prime_r,  x^{\prime \prime}_r \right)$ is a pair of dummy variables. Then, insertion into~\eqref{Xstar} of~\eqref{CYHN}--\eqref{CHN} (with~$\xi_3 = 0$) yields:
\begin{equation} \label{43}
\frac{X_{rr} \left( R \right)}{\sigma^2_Y} = \frac{R}{3} \int^R_0 \dd u  \left( 2 - 3 \frac{u}{R} + \frac{u^3}{R^3} \right) \rho_Y \left( u \right) + \frac{\sqrt{2/\pi}}{R^2} \, \bar{\mathcal{X}}_\star \left( R \right),
\end{equation}
where we have set:
\begin{equation} \label{A14}
\bar{\mathcal{X}}_\star = \int^\infty_0 \int^\infty_0 \int^R_0 \int^R_0 \dd k_r \, \dd k_3 \, \dd x \, \dd y \,  k_r \, \tilde \rho_Y \left( k_r, k_3 \right) y^2 \frac{\partial}{\partial y} \left [x^2 \frac{\partial}{\partial x} \Psi_H \left( x, y \right) - x \,\Psi_{YH} \left( x, y \right) \right ],
\end{equation}
\begin{equation}
\Psi_{YH} \left( x, y \right) = J_0 \left( k_r \left| x - y \right| \right) \mathrm{K}_0 (k y) - J_0 (k_r x) \, \mathrm{K}_0 (k_3 y), \quad k = \sqrt{k^2_r + k^2_3} \,,
\end{equation}
\begin{equation}
\Psi_H \left( x, y \right) = \mathrm{K}_0 (k x) \, \Psi_{YH} \left( x, y \right) + \mathrm{K}_0 (k_3 x) \left [ \mathrm{K}_0 (k_3 y) - J_0 \left(k_r y \right) \mathrm{K}_0 (k y) \right].
\end{equation}
Hence, integration by parts in the domain~$\left[ 0, R \right] \times \left[ 0, R \right]$ enables one to decompose the integral~\eqref{A14} as~$\bar{\mathcal{X}}_\star = 4 \mathcal{X}_4 - 2 R^2 \mathcal{X}_3 + R^4 \mathcal{X}_2$, with
\begin{equation}
\mathcal{X}_2 \left( R \right) = \int^\infty_0 \int^\infty_0 \dd k_r \, \dd k_3 \, k_r \, \tilde \rho_Y \left( k_r, k_3 \right) \Psi_H \left( R, R \right),
\end{equation}
\begin{equation}
\mathcal{X}_3 \left( R \right) = \int^\infty_0 \int^\infty_0 \int^R_0 \dd k_r \, \dd k_3 \, \dd x \, k_r \, \tilde \rho_Y \left( k_r, k_3 \right) x \left[ \Psi_H \left( x, R \right) + \frac{1}{2} \Psi_{YH} \left( x, R \right) + \Psi_H \left( R, x \right) \right],
\end{equation}
\begin{equation}
\mathcal{X}_4 \left( R \right) = \int^\infty_0 \int^\infty_0 \int^R_0 \int^R_0 \dd k_r \, \dd k_3 \, \dd x \, \dd y \, k_r \, \tilde \rho_Y \left( k_r, k_3 \right) x \, y \left[ \Psi_H \left( x, y \right) + \frac{1}{2} \, \Psi_{YH} \left( x, y \right) \right].
\end{equation}

The utility related to the decomposition in~\eqref{dec}, and the subsequent developments, relies on the
fact that one can clearly distinguish the contribution (\ie~$\mathcal{X}_\infty$) due to the mean radial flow from that (\ie~$\mathcal{X}_\star$) associated to the fluctuation of the head-gradient.
\begin{figure}
\includegraphics[width=1.0\textwidth]{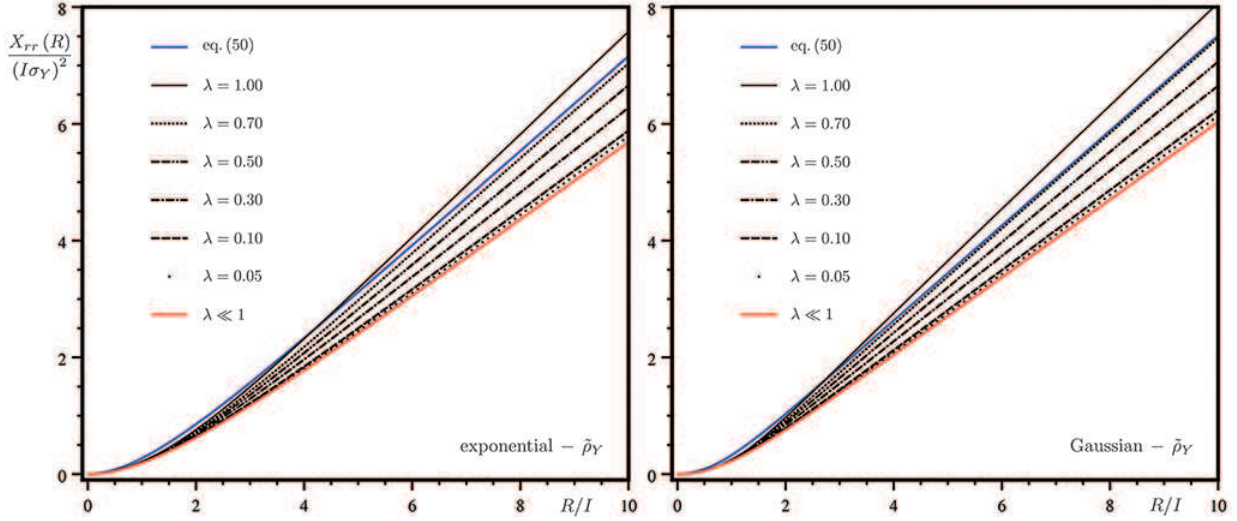}
\vspace{-20 pt}
\caption{Scaled trajectory variance~$X_{rr} / \left( I \sigma_Y \right)^2$ as computed from~\eqref{43} for several values of the anisotropy ratio~$\lambda$ (exponential and Gaussian spectrum~$\tilde{\rho}_Y$). Continuous red and blue lines refer to eqs~\eqref{PG} and~\eqref{mine}, respectively.}
\label{Figure5}
\end{figure}
In the FIG.~\ref{Figure5} we have depicted the scaled moment~$X_{rr} / \left( I \sigma_Y \right)^2$ versus the non dimensional travel distance~$R / I$. It has been done for both exponential and Gaussian spectrum.  For comparison purposes, we have also depicted (continuous red line) the approximation~$X_{rr} \simeq \mathcal{X}_\infty$:
\begin{equation}
X_{rr} \left( R \right) \simeq \frac{\left( I \sigma_Y \right)^2}{3 \pi^2 R^2} \begin{cases}
\pi^2 \left [2 R^3 - 3 R^2 +6 - 6 \left( R +1 \right) \exp \left( -R \right) \right] & \mathrm{(exp)} \\
8 - 6 \pi R^2 + 2 \pi^2 R^3 \mathrm{erf} \left( \displaystyle \frac{\sqrt \pi}{2} R \right) + 4 \left( \pi R^2 - 2 \right) \exp \left( - \displaystyle \frac{\pi}{4} \, R^2 \right) & \mathrm{(Gauss)}
\end{cases}
\label{PG}
\end{equation}
\citep{Indelman1999} along with (continuous blue line) a newly derived (for details, see the APPENDIX) approximate expression of~$X_{rr}$, \ie
\begin{equation}
X_{rr} \simeq \frac{\left( I \sigma_Y \right)^2}{27 \pi^2 R^2} \begin{cases}
\pi^2 \left[ 22 R^3 - 27 R^2 +30 + 6 \left( 2 R^2 - 5 R - 5 \right) \exp \left( -R \right) \right] & \mathrm{(exp)}\\
40 - 54 \pi R^2 + 22 \pi^2 R^3 \mathrm{erf} \left( \displaystyle \frac{\sqrt \pi}{2} R \right) + 4 \left( 11 \pi R^2 - 10 \right) \exp \left( - \displaystyle \frac{\pi}{4} \, R^2 \right) & \mathrm{(Gauss).}
\end{cases}
\label{mine}
\end{equation}

As particles are injected through the source in the porous medium, the radial moment~$X_{rr}$ increases monotonically with~$R$. At short distances,~$X_{rr}$
displays a nonlinear dependence, whereas at large distances it grows linearly. These findings rely upon the dependence of~$X_{rr}$ on the velocity covariance through eq.~\eqref{Xrrr} that, in turn, is a measure of the distance over which the velocities of two fluid particles are correlated. As a consequence, for~$R \ll I$ two fluid particles have not covered a single integral scale~$I$, and concurrently they are highly correlated. As a consequence, scattering results enhanced by the dominant impact of the velocity covariance~$u_{rr}$. Conversely, at large distances the advective velocity drops like~$x^{-1}_r$, and the net, overall effect is still an increasing scattering, but with a lesser gradient. In order to address such a behavior in a quantitative manner, one can refer either to the approximate expression of~\cite{Indelman1999}, \ie~$X_{rr} \left( R \right) \simeq \left( \sigma^2_Y R / 3 \right) \int^R_0 \dd u \left( 2 - 3 u / R + u^3 / R^3 \right) \rho_Y \left( u \right)$, or to eq.~\eqref{A13}, \ie~$X_{rr} \left( R \right) \simeq \left (\sigma^2_Y R / 27 \right) \int^R_0 \dd u  \left( 22 - 27 u / R + 5 u^3 / R^3 \right) \rho_Y \left( u \right)$. Thus, at small distances it yields~$\rho_Y \sim 1$, and one recovers that~$X_{rr} \sim R^2$. Instead, at large~$R$ one has~$u/R = \mathrm{o} \left( 1 \right)$, and therefore~$X_{rr} \sim R \int^\infty_0 \dd u \, \rho_Y \left( u \right) = R$. The reduction of~$X_{rr}$ with the small~$\lambda$-values is explained similarly to the above discussion: for a solute particles it is easier to circumvent, by taking a vertical step, a poorly conducting inclusion characterized by~$\lambda \ll 1$ as compared with an inclusion of quasi isotropic (\ie~$\lambda \simeq 1$) heterogeneity's structure. As a consequence, the deviation from the mean is larger in the latter case, and this explains the increasing (for given~$R$) trajectory's variance as~$\lambda \to 1$. Finally, it is seen that the approximation~\eqref{PG} is found in excellent agreement with the complete solution in the case of strongly anisotropic ($\lambda \ll 1$) formations. It also provides a lower bound for~$X_{rr}$, since it neglects part (the one associated to the mean gradient) of the scattering process. Instead, the expression~\eqref{mine} is found in a reasonable agreement in the other regime of pseudo-isotropic ($\lambda \lesssim 1$) formations. Equations~\eqref{PG}--\eqref{mine} are straightforwardly extended to disordered media of axial symmetric
heterogeneity's structure by replacing~$R \to R / \sqrt{\cos^2 \phi + \lambda^{-2} \sin^2 \phi} \,$, being~$\phi$ the angle between the mean trajectory and the plane of isotropy. 

\section*{Concluding remarks}
Scattering processes generated by localized/distributed sources are a powerful tool which finds application in numerous branches of applied sciences. In quantum physics, scattering is used to infer the size as well as the distribution of the electrical charge of nuclei, whereas in the electrodynamics it serves to compute dielectric properties. In the theory of composites and in the reservoir engineering (the fields of main concern for the present study), it serves to identify the effective (flow and transport) properties of disordered media. \\
\indent We have focused on scattering of a passive scalar injected in a formation and advected by a steady velocity, that in turn is generated by a line of singularity. Within a homogeneous domain, the solute propagates by advection like
a cylinder of radius~$R \equiv R \left( t \right)$, whereas scattering is due to the diffusion mechanism, solely. In disordered media, scattering is determined by the fluctuations of the advective velocity which are caused by the erratic, spatial variability of the conductivity~$K$. Within a stochastic framework, that regards the log-conductivity~$\ln K$ as a stationary, Gaussian, random field, scattering is quantified by means of the second-order radial moment which, by virtue of ergodicity, coincides with the trajectory variance~\eqref{Xrrr}. After adopting a few simplifying assumptions (the most relevant of which requires that the variance of~$\ln K$ is much smaller than one), it is shown that, central for the study, is the computation of the integral~\eqref{model}. Despite its origin, it is recognized that such a quantity is instrumental for many other problems arising in several branches of classical as well as quantum physics, and therefore its study results of a much wider interest than that strictly considered here. The analytical computation of~\eqref{model} is achieved as large time limit of the same problem in the unsteady state flow regime. \\
\indent Unlike past studies on the same topic~\citep[see, \eg][]{Fiori1998}, here covariances of the flow variables are expressed in terms of two quadratures solely, which are easily carried out after specifying the shape of the spectrum (the Fourier transform of the autocorrelation of~$Y$). Illustrations focus on the (cross)-variances of the specific energy and the radial velocity, since they are usually of interest in the applications. It is seen that, although the log-conductivity
is a stationary random field, these variances are not since the mean flow is not uniform. \\
\indent The trajectory variance~$X_{rr}$ is computed and discussed for both exponential and Gaussian spectrum, being these models generally adopted in the real world applications~\citep{dagan1989}. In particular, the transitional regime from the early to the large distances is much more persistent than that pertaining to the approximation valid for formations with an anisotropic ratio~$\lambda$ much lesser than one~\citep{Indelman1999}.  This is due to the impact of the covariances involving
the head-gradient, which in their approximation are neglected. Conversely, it is achieved a new, analytical expression of~$X_{rr}$ that accounts, reasonably well, for scattering in formations with~$\lambda$ close to~$1$.

\vspace{0.25 cm}

\begin{acknowledgments}
The present study was developed within the \textsf{GNCS} (\emph{Gruppo Nazionale Calcolo Scientifico} - INdAM) framework, and it was supported by the project $\# \, 3778/2022$ (Departmental fund). The final release of all the figures was achieved thanks to the computer artistry of Dr Gugliemo~\textsf{BRUNETTI}, to whom we are greatly indebted.
\end{acknowledgments}

\newpage

\appendix*
\noindent \textbf{APPENDIX: derivation of the approximate expression~\eqref{mine}} \\
\indent As a preparatory step, we re-write the last of~\eqref{Xstar}  as:

\begin{align}
& \mathcal{X}_\star \left( R \right) = \left( \frac{2 \pi}{QR} \right)^2 \int^R_0 \int^R_0 \dd x^\prime_r \, \dd x^{\prime \prime}_r \left( x^\prime_r \, x^{\prime \prime}_r \right)^2 \frac{\partial}{\partial x^{\prime \prime}_r} \left [ \frac{\partial}{\partial x^\prime_r} \, C_H \left( x^\prime_r, x^{\prime \prime}_r \right) - \frac{Q}{\pi x^\prime_r} \, C_{YH} \left( x^\prime_r, x^{\prime \prime}_r \right) \right ] = \notag \\
& \left( \frac{2 \pi}{QR} \right)^2 \int^R_0 \int^R_0 \dd x^\prime_r \, \dd x^{\prime \prime}_r \left( x^\prime_r \, x^{\prime \prime}_r \right)^2 \frac{\partial}{\partial x^{\prime \prime}_r} \left [ \frac{\partial \, C_H \left( x^\prime_r, x^{\prime \prime}_r \right)}{\partial x^\prime_r} + 2 \, C_{YH} \left( x^\prime_r, x^{\prime \prime}_r \right) \frac{\partial \, H^{(0)} \left( x^\prime_r \right)}{\partial x^\prime_r} \right ].
\tag{A1} \label{A1}
\end{align}
Then, the last double integral in~\eqref{A1} is re-written as:

\begin{align}
& \int^R_0 \dd x^{\prime \prime}_r \, x^{\prime \prime \, 2}_r \frac{\partial}{\partial x^{\prime \prime}_r} \int^R_0 \dd x^\prime_r \, x^{\prime \, 2}_r \left [ \frac{\partial \, C_H \left( x^\prime_r, x^{\prime \prime}_r \right)}{\partial x^\prime_r} + 2 \, C_{YH} \left( x^\prime_r, x^{\prime \prime}_r \right) \frac{\partial \, H^{(0)} \left( x^\prime_r \right)}{\partial x^\prime_r} \right ] \simeq \notag \\
& - \frac{1}{3} \int^R_0 \int^R_0 \dd x^\prime_r \, \dd x^{\prime \prime}_r \, x^{\prime \, 3}_r x^{\prime \prime \, 2}_r \frac{\partial}{\partial x^{\prime \prime}_r} \left [ \frac{\partial^2 \, C_H \left( x^\prime_r, x^{\prime \prime}_r \right)}{\partial x^{\prime \, 2}_r} + 2 \frac{\partial \, H^{(0)} \left( x^\prime_r \right)}{\partial x^\prime_r} \frac{\partial \, C_{YH} \left( x^\prime_r, x^{\prime \prime}_r \right)}{\partial x^\prime_r} \right ],
\tag{A2} \label{A2}
\end{align}
where the second passage in~\eqref{A2} has been achieved upon integration by parts and neglecting the finite term due to its very fast (exponential) decay with~$R$. In addition, the term~$2 x^{\prime \, 3}_r C_{YH} \left( x^\prime_r, x^{\prime \prime}_r \right) \frac{\partial^2 }{\partial x^2_r} H^{(0)} \left( x_r \right)$ (that also arises upon application of integration by parts) has been dropped out, since, from the definition of two-dimensional Green function, one has~$\frac{\partial^2 }{\partial x^2_r} H^{(0)} \left( x_r \right) = - Q \delta \left( \bm{x}_r \right)$. \\
\indent At this stage, we note that the governing equation~\eqref{flowes} for the head's fluctuation can be written in approximate manner as follows:
\begin{equation}
- \left ( \nabla^2_r + \frac{\partial^2}{\partial x^2_3} \right) H^{(1)} \left( \bm x \right) \simeq - \nabla^2_r H^{(1)} \left( \bm x \right) = \nabla_r \, H^{(0)} \left( x_r \right) \cdot \nabla_r Y \left( \bm x \right).
\tag{A3} \label{A3}
\end{equation}
The neglect of the second-order derivative~$\frac{\partial^2}{\partial x^2_3}$ as compared with the laplacian~$\nabla^2_r$ is authorized by the fact that most of the flow develops radially, and therefore the dominant variations of the head's fluctuation occur in the horizontal plane. In order to provide a quantitative reasoning, we recall that~$\frac{\partial^2}{\partial x^2_3} \sim \mathcal{O} \left( I^{-2}_v \right)$, whereas~$\nabla^2_r \sim \mathcal{O} \left( I^{-2} \right)$. As a consequence, the ratio of the two estimates behaves like~$\left( I_v / I \right)^2 = \lambda^2$. Since, the majority of the natural formations are anisotropic ($\lambda \le 1$), we argue that the above approximation works quite well~\citep[see also discussion in][]{Indelman1999}. Hence, upon multiplication of~\eqref{A3} by the head's fluctuation evaluated at~$\bm{y}_r \ne \bm{x}_r$, and taking the ensemble average, it leads to:
\begin{equation}
- \nabla^2_r \, C_H \left( x_r, y_r \right) = \nabla_r \, H^{(0)} \left( x_r \right) \cdot \nabla_r \, C_{YH} \left( x_r, y_r \right).
\tag{A4} \label{A4}
\end{equation}
Then, application of the chain rule of derivation~$\frac{\partial}{\partial x_m} \equiv \frac{x_m}{x_r} \frac{\partial}{\partial x_r} \, \, (m=1,2)$ enables one to write~\eqref{A4} as:
\begin{equation}
\frac{\partial^2}{\partial x^2_r} \, C_H \left( x_r, y_r \right) = - \frac{\partial}{\partial x_r} \, H^{(0)} \left( x_r \right) \frac{\partial}{\partial x_r} \, C_{YH} \left( x_r, y_r \right),
\tag{A5}
\end{equation}
and the subsequent substitution into the last of~\eqref{A2} permits to write~$\mathcal{X}_\star$ as:
\begin{equation}
\mathcal{X}_\star \left( R \right) = - \frac{1}{3} \left( \frac{2 \pi}{QR} \right)^2 \int^R_0 \int^R_0 \dd x^\prime_r \, \dd x^{\prime \prime}_r  \, x^{\prime \, 3}_r x^{\prime \prime \, 2}_r \frac{\partial \, H^{(0)} \left( x^\prime_r \right)}{\partial x^\prime_r} \frac{\partial^2 \, C_{YH} \left( x^\prime_r, x^{\prime \prime}_r \right)}{\partial x^\prime_r \partial x^{\prime \prime}_r}.
\tag{A6} \label{A6}
\end{equation}
By taking integration by parts in~\eqref{A6} with respect to the variable~$x^{\prime \prime}_r$, it yields (with the same reasoning as before):
\begin{equation}
\mathcal{X}_\star \left( R \right) = \left( \frac{2 \pi}{3QR} \right)^2 \int^R_0 \int^R_0 \dd x^\prime_r \, \dd x^{\prime \prime}_r  \left ( x^\prime_r x^{\prime \prime}_r \right)^3 \frac{\partial \, H^{(0)} \left( x^\prime_r \right)}{\partial x^\prime_r} \frac{\partial}{\partial x^\prime_r} \left [ \frac{\partial^2 \, C_{YH} \left( x^\prime_r, x^{\prime \prime}_r \right)}{\partial x^{\prime \prime \, 2}_r} \right].
\tag{A7} \label{A7}
\end{equation}
Likewise, one can write:
\begin{equation}
\frac{\partial^2}{\partial y^2_r} \, C_{YH} \left( x_r, y_r \right) = - \sigma^2_Y \frac{\partial}{\partial y_r} H^{(0)} \left( y_r \right) \frac{\partial}{\partial y_r} \rho_Y \left( x_r - y_r \right),
\tag{A8}
\end{equation}
and therefore eq.~\eqref{A7} reads as:
\begin{equation}
\mathcal{X}_\star \left( R \right) = - \left( \frac{2 \pi \sigma_Y}{3QR} \right)^2 \int^R_0 \int^R_0 \dd x^\prime_r \, \dd x^{\prime \prime}_r  \left ( x^\prime_r x^{\prime \prime}_r \right)^3 \frac{\partial \, H^{(0)} \left( x^\prime_r \right)}{\partial x^\prime_r} \frac{\partial^2 \, \rho_Y \left( x^\prime_r - x^{\prime \prime}_r \right)}{\partial x^\prime_r \partial x^{\prime \prime}_r} \frac{\partial \, H^{(0)} \left( x^{\prime \prime}_r \right)}{\partial x^{\prime \prime}_r}.
\tag{A9} \label{A9}
\end{equation}
By noting that:
\begin{equation}
\frac{\partial}{\partial x_r} H^{(0)} \left( x_r \right) = - \frac{Q}{2 \pi x_r}, \qquad \frac{\partial^2 }{\partial x_r \partial y_r} \, \rho_Y \left( x_r - y_r \right) \equiv - \frac{\dd^2}{\dd u^2} \, \rho_Y \left( u \right) \bigg |_{u = x_r - y_r},
\tag{A10}
\end{equation}
eq.~\eqref{A9} becomes:
\begin{equation}
\mathcal{X}_\star \left( R \right) = \left( \frac{\sigma_Y}{3R} \right)^2 \int^R_0 \int^R_0 \dd x^\prime_r \, \dd x^{\prime \prime}_r  \left ( x^\prime_r x^{\prime \prime}_r \right)^2 \frac{\dd^2}{\dd u^2} \, \rho_Y \left( u \right) \bigg |_{u = x^\prime_r - x^{\prime \prime}_r}.
\tag{A11}
\end{equation}
Hence, the computation of one quadrature leads to:
\begin{equation}
\mathcal{X}_\star \left( R \right) = \frac{\sigma^2_Y}{135} \, R^3 \int^R_0 \dd u \left( 6 - 15 \frac{u}{R} + 10 \frac{u^2}{R^2} - \frac{u^5}{R^5} \right) \frac{\dd^2}{\dd u^2} \, \rho_Y \left( u \right),
\tag{A12} \label{A12}
\end{equation}
and the application (two times) of integration by parts provides (on the same grounds of the above adopted approximation) the final result:
\begin{equation} \tag{A13} \label{A13}
X_{rr} \left( R \right) = \mathcal{X}_\infty \left( R \right) + \mathcal{X}_\star \left( R \right) \simeq \frac{\sigma^2_Y}{27} \, R \int^R_0 \dd u  \left( 22 - 27 \frac{u}{R} + 5 \frac{u^3}{R^3} \right) \rho_Y \left( u \right).
\end{equation}
Finally, insertion into~\eqref{A13} of exponential and Gaussian autocorrelation~$\rho_Y$ leads to~\eqref{mine}.

\newpage

\nocite{*}
\bibliography{scattering}

\begin{thebibliography}{26}%
\makeatletter
\providecommand \@ifxundefined [1]{%
 \@ifx{#1\undefined}
}%
\providecommand \@ifnum [1]{%
 \ifnum #1\expandafter \@firstoftwo
 \else \expandafter \@secondoftwo
 \fi
}%
\providecommand \@ifx [1]{%
 \ifx #1\expandafter \@firstoftwo
 \else \expandafter \@secondoftwo
 \fi
}%
\providecommand \natexlab [1]{#1}%
\providecommand \enquote  [1]{``#1''}%
\providecommand \bibnamefont  [1]{#1}%
\providecommand \bibfnamefont [1]{#1}%
\providecommand \citenamefont [1]{#1}%
\providecommand \href@noop [0]{\@secondoftwo}%
\providecommand \href [0]{\begingroup \@sanitize@url \@href}%
\providecommand \@href[1]{\@@startlink{#1}\@@href}%
\providecommand \@@href[1]{\endgroup#1\@@endlink}%
\providecommand \@sanitize@url [0]{\catcode `\\12\catcode `\$12\catcode
  `\&12\catcode `\#12\catcode `\^12\catcode `\_12\catcode `\%12\relax}%
\providecommand \@@startlink[1]{}%
\providecommand \@@endlink[0]{}%
\providecommand \url  [0]{\begingroup\@sanitize@url \@url }%
\providecommand \@url [1]{\endgroup\@href {#1}{\urlprefix }}%
\providecommand \urlprefix  [0]{URL }%
\providecommand \Eprint [0]{\href }%
\providecommand \doibase [0]{http://dx.doi.org/}%
\providecommand \selectlanguage [0]{\@gobble}%
\providecommand \bibinfo  [0]{\@secondoftwo}%
\providecommand \bibfield  [0]{\@secondoftwo}%
\providecommand \translation [1]{[#1]}%
\providecommand \BibitemOpen [0]{}%
\providecommand \bibitemStop [0]{}%
\providecommand \bibitemNoStop [0]{.\EOS\space}%
\providecommand \EOS [0]{\spacefactor3000\relax}%
\providecommand \BibitemShut  [1]{\csname bibitem#1\endcsname}%
\let\auto@bib@innerbib\@empty
\bibitem [{\citenamefont {Abramovich}\ and\ \citenamefont
  {Indelman}(1995)}]{abramovich1995}%
  \BibitemOpen
  \bibfield  {author} {\bibinfo {author} {\bibnamefont {Abramovich},
  \bibfnamefont {B.}}\ and\ \bibinfo {author} {\bibnamefont {Indelman},
  \bibfnamefont {P.}},\ }\bibfield  {title} {\enquote {\bibinfo {title}
  {Effective permittivity of log-normal isotropic random media},}\ }\href@noop
  {} {\bibfield  {journal} {\bibinfo  {journal} {Journal of Physics A:
  Mathematical and General}\ }\textbf {\bibinfo {volume} {28}},\ \bibinfo
  {pages} {693--700} (\bibinfo {year} {1995})}\BibitemShut {NoStop}%
\bibitem [{\citenamefont {Bahraminasab}\ \emph {et~al.}(2008)\citenamefont
  {Bahraminasab}, \citenamefont {Niry}, \citenamefont {Davoudi}, \citenamefont
  {Reza Rahimi~Tabar}, \citenamefont {Masoudi},\ and\ \citenamefont
  {Sreenivasan}}]{Bahraminasab08}%
  \BibitemOpen
  \bibfield  {author} {\bibinfo {author} {\bibnamefont {Bahraminasab},
  \bibfnamefont {A.}}, \bibinfo {author} {\bibnamefont {Niry}, \bibfnamefont
  {M.~D.}}, \bibinfo {author} {\bibnamefont {Davoudi}, \bibfnamefont {J.}},
  \bibinfo {author} {\bibnamefont {Reza Rahimi~Tabar}, \bibfnamefont {M.}},
  \bibinfo {author} {\bibnamefont {Masoudi}, \bibfnamefont {A.~A.}}, \ and\
  \bibinfo {author} {\bibnamefont {Sreenivasan}, \bibfnamefont {K.~R.}},\
  }\bibfield  {title} {\enquote {\bibinfo {title} {Taylor's frozen-flow
  hypothesis in burgers turbulence},}\ }\href {\doibase
  10.1103/PhysRevE.77.065302} {\bibfield  {journal} {\bibinfo  {journal} {Phys.
  Rev. E}\ }\textbf {\bibinfo {volume} {77}},\ \bibinfo {pages} {065302}
  (\bibinfo {year} {2008})}\BibitemShut {NoStop}%
\bibitem [{\citenamefont {Blanco-Canosa}\ \emph {et~al.}(2014)\citenamefont
  {Blanco-Canosa}, \citenamefont {Frano}, \citenamefont {Schierle},
  \citenamefont {Porras}, \citenamefont {Loew}, \citenamefont {Minola},
  \citenamefont {Bluschke}, \citenamefont {Weschke}, \citenamefont {Keimer},\
  and\ \citenamefont {Le~Tacon}}]{blanco2014}%
  \BibitemOpen
  \bibfield  {author} {\bibinfo {author} {\bibnamefont {Blanco-Canosa},
  \bibfnamefont {S.}}, \bibinfo {author} {\bibnamefont {Frano}, \bibfnamefont
  {A.}}, \bibinfo {author} {\bibnamefont {Schierle}, \bibfnamefont {E.}},
  \bibinfo {author} {\bibnamefont {Porras}, \bibfnamefont {J.}}, \bibinfo
  {author} {\bibnamefont {Loew}, \bibfnamefont {T.}}, \bibinfo {author}
  {\bibnamefont {Minola}, \bibfnamefont {M.}}, \bibinfo {author} {\bibnamefont
  {Bluschke}, \bibfnamefont {M.}}, \bibinfo {author} {\bibnamefont {Weschke},
  \bibfnamefont {E.}}, \bibinfo {author} {\bibnamefont {Keimer}, \bibfnamefont
  {B.}}, \ and\ \bibinfo {author} {\bibnamefont {Le~Tacon}, \bibfnamefont
  {M.}},\ }\bibfield  {title} {\enquote {\bibinfo {title} {Resonant x-ray
  scattering study of charge-density wave correlations
  in~$\mathrm{YBa}_2\mathrm{Cu}_3\mathrm{O}_{6+x}$},}\ }\href@noop {}
  {\bibfield  {journal} {\bibinfo  {journal} {Physical Review B}\ }\textbf
  {\bibinfo {volume} {90}},\ \bibinfo {pages} {054513} (\bibinfo {year}
  {2014})}\BibitemShut {NoStop}%
\bibitem [{\citenamefont {Chin}(1997)}]{chin1997}%
  \BibitemOpen
  \bibfield  {author} {\bibinfo {author} {\bibnamefont {Chin}, \bibfnamefont
  {D.~A.}},\ }\bibfield  {title} {\enquote {\bibinfo {title} {An assessment of
  first-order stochastic dispersion theories in porous media},}\ }\href@noop {}
  {\bibfield  {journal} {\bibinfo  {journal} {Journal of hydrology}\ }\textbf
  {\bibinfo {volume} {199}},\ \bibinfo {pages} {53--73} (\bibinfo {year}
  {1997})}\BibitemShut {NoStop}%
\bibitem [{\citenamefont {Dagan}(1989)}]{dagan1989}%
  \BibitemOpen
  \bibfield  {author} {\bibinfo {author} {\bibnamefont {Dagan}, \bibfnamefont
  {G.}},\ }\href@noop {} {\emph {\bibinfo {title} {Flow and Transport in Porous
  Formation}}}\ (\bibinfo  {publisher} {Springer-Verlag},\ \bibinfo {address}
  {New York},\ \bibinfo {year} {1989})\BibitemShut {NoStop}%
\bibitem [{\citenamefont {Fiori}, \citenamefont {Indelman},\ and\ \citenamefont
  {Dagan}(1998)}]{Fiori1998}%
  \BibitemOpen
  \bibfield  {author} {\bibinfo {author} {\bibnamefont {Fiori}, \bibfnamefont
  {A.}}, \bibinfo {author} {\bibnamefont {Indelman}, \bibfnamefont {P.}}, \
  and\ \bibinfo {author} {\bibnamefont {Dagan}, \bibfnamefont {G.}},\
  }\bibfield  {title} {\enquote {\bibinfo {title} {Correlation structure of
  flow variables for steady flow toward a well with application to highly
  anisotropic heterogeneous formations},}\ }\href {\doibase 10.1029/97WR02491}
  {\bibfield  {journal} {\bibinfo  {journal} {Water Resources Research}\
  }\textbf {\bibinfo {volume} {34}},\ \bibinfo {pages} {699--708} (\bibinfo
  {year} {1998})}\BibitemShut {NoStop}%
\bibitem [{\citenamefont {Gradshteyn}\ and\ \citenamefont
  {Ryzhik}(2014)}]{gradshteyn2014table}%
  \BibitemOpen
  \bibfield  {author} {\bibinfo {author} {\bibnamefont {Gradshteyn},
  \bibfnamefont {I.~S.}}\ and\ \bibinfo {author} {\bibnamefont {Ryzhik},
  \bibfnamefont {I.~M.}},\ }\href@noop {} {\emph {\bibinfo {title} {Table of
  integrals, series, and products}}}\ (\bibinfo  {publisher} {Academic press},\
  \bibinfo {year} {2014})\BibitemShut {NoStop}%
\bibitem [{\citenamefont {Indelman}(1996)}]{indelman1996averaging}%
  \BibitemOpen
  \bibfield  {author} {\bibinfo {author} {\bibnamefont {Indelman},
  \bibfnamefont {P.}},\ }\bibfield  {title} {\enquote {\bibinfo {title}
  {Averaging of unsteady flows in heterogeneous media of stationary
  conductivity},}\ }\href@noop {} {\bibfield  {journal} {\bibinfo  {journal}
  {Journal of Fluid Mechanics}\ }\textbf {\bibinfo {volume} {310}},\ \bibinfo
  {pages} {39--60} (\bibinfo {year} {1996})}\BibitemShut {NoStop}%
\bibitem [{\citenamefont {Indelman}\ and\ \citenamefont
  {Dagan}(1999)}]{Indelman1999}%
  \BibitemOpen
  \bibfield  {author} {\bibinfo {author} {\bibnamefont {Indelman},
  \bibfnamefont {P.}}\ and\ \bibinfo {author} {\bibnamefont {Dagan},
  \bibfnamefont {G.}},\ }\bibfield  {title} {\enquote {\bibinfo {title} {Solute
  transport in divergent radial flow through heterogeneous porous media},}\
  }\href@noop {} {\bibfield  {journal} {\bibinfo  {journal} {Journal of Fluid
  Mechanics}\ }\textbf {\bibinfo {volume} {384}},\ \bibinfo {pages} {159--182}
  (\bibinfo {year} {1999})}\BibitemShut {NoStop}%
\bibitem [{\citenamefont {Indelman}\ and\ \citenamefont
  {Rubin}(1996)}]{Indelman1996}%
  \BibitemOpen
  \bibfield  {author} {\bibinfo {author} {\bibnamefont {Indelman},
  \bibfnamefont {P.}}\ and\ \bibinfo {author} {\bibnamefont {Rubin},
  \bibfnamefont {Y.}},\ }\bibfield  {title} {\enquote {\bibinfo {title} {Solute
  transport in nonstationary velocity fields},}\ }\href@noop {} {\bibfield
  {journal} {\bibinfo  {journal} {Water resources research}\ }\textbf {\bibinfo
  {volume} {32}},\ \bibinfo {pages} {1259--1267} (\bibinfo {year}
  {1996})}\BibitemShut {NoStop}%
\bibitem [{\citenamefont {Jackson}(2007)}]{jackson2007}%
  \BibitemOpen
  \bibfield  {author} {\bibinfo {author} {\bibnamefont {Jackson}, \bibfnamefont
  {J.~D.}},\ }\href@noop {} {\emph {\bibinfo {title} {Classical
  electrodynamics}}}\ (\bibinfo  {publisher} {John Wiley \& Sons},\ \bibinfo
  {address} {New York},\ \bibinfo {year} {2007})\BibitemShut {NoStop}%
\bibitem [{\citenamefont {Koplik}, \citenamefont {Redner},\ and\ \citenamefont
  {Hinch}(1994)}]{koplik94}%
  \BibitemOpen
  \bibfield  {author} {\bibinfo {author} {\bibnamefont {Koplik}, \bibfnamefont
  {J.}}, \bibinfo {author} {\bibnamefont {Redner}, \bibfnamefont {S.}}, \ and\
  \bibinfo {author} {\bibnamefont {Hinch}, \bibfnamefont {E.}},\ }\bibfield
  {title} {\enquote {\bibinfo {title} {Tracer dispersion in planar multipole
  flows},}\ }\href {\doibase 10.1103/PhysRevE.50.4650} {\bibfield  {journal}
  {\bibinfo  {journal} {Physical Review E}\ }\textbf {\bibinfo {volume} {50}},\
  \bibinfo {pages} {4650} (\bibinfo {year} {1994})}\BibitemShut {NoStop}%
\bibitem [{\citenamefont {Kurowski}\ \emph {et~al.}(1994)\citenamefont
  {Kurowski}, \citenamefont {Ippolito}, \citenamefont {Hulin}, \citenamefont
  {Koplik},\ and\ \citenamefont {Hinch}}]{kurowski1994}%
  \BibitemOpen
  \bibfield  {author} {\bibinfo {author} {\bibnamefont {Kurowski},
  \bibfnamefont {P.}}, \bibinfo {author} {\bibnamefont {Ippolito},
  \bibfnamefont {I.}}, \bibinfo {author} {\bibnamefont {Hulin}, \bibfnamefont
  {J.}}, \bibinfo {author} {\bibnamefont {Koplik}, \bibfnamefont {J.}}, \ and\
  \bibinfo {author} {\bibnamefont {Hinch}, \bibfnamefont {E.}},\ }\bibfield
  {title} {\enquote {\bibinfo {title} {Anomalous dispersion in a dipole flow
  geometry},}\ }\href {\doibase 10.1063/1.868075} {\bibfield  {journal}
  {\bibinfo  {journal} {Physics of Fluids}\ }\textbf {\bibinfo {volume} {6}},\
  \bibinfo {pages} {108--117} (\bibinfo {year} {1994})}\BibitemShut {NoStop}%
\bibitem [{\citenamefont {Le~Borgne}, \citenamefont {Dentz},\ and\
  \citenamefont {Carrera}(2008)}]{leborgne08}%
  \BibitemOpen
  \bibfield  {author} {\bibinfo {author} {\bibnamefont {Le~Borgne},
  \bibfnamefont {T.}}, \bibinfo {author} {\bibnamefont {Dentz}, \bibfnamefont
  {M.}}, \ and\ \bibinfo {author} {\bibnamefont {Carrera}, \bibfnamefont
  {J.}},\ }\bibfield  {title} {\enquote {\bibinfo {title} {Lagrangian
  statistical model for transport in highly heterogeneous velocity fields},}\
  }\href {\doibase 10.1103/PhysRevLett.101.090601} {\bibfield  {journal}
  {\bibinfo  {journal} {Phys. Rev. Lett.}\ }\textbf {\bibinfo {volume} {101}},\
  \bibinfo {pages} {090601} (\bibinfo {year} {2008})}\BibitemShut {NoStop}%
\bibitem [{\citenamefont {Martin}\ and\ \citenamefont
  {Shaw}(2019)}]{martin2019}%
  \BibitemOpen
  \bibfield  {author} {\bibinfo {author} {\bibnamefont {Martin}, \bibfnamefont
  {B.~R.}}\ and\ \bibinfo {author} {\bibnamefont {Shaw}, \bibfnamefont {G.}},\
  }\href@noop {} {\emph {\bibinfo {title} {Nuclear and particle physics: an
  introduction}}}\ (\bibinfo  {publisher} {John Wiley \& Sons},\ \bibinfo
  {year} {2019})\BibitemShut {NoStop}%
\bibitem [{\citenamefont {Renau}, \citenamefont {Read},\ and\ \citenamefont
  {Brunt}(1982)}]{renau1982}%
  \BibitemOpen
  \bibfield  {author} {\bibinfo {author} {\bibnamefont {Renau}, \bibfnamefont
  {A.}}, \bibinfo {author} {\bibnamefont {Read}, \bibfnamefont {F.}}, \ and\
  \bibinfo {author} {\bibnamefont {Brunt}, \bibfnamefont {J.}},\ }\bibfield
  {title} {\enquote {\bibinfo {title} {The charge-density method of solving
  electrostatic problems with and without the inclusion of space-charge},}\
  }\href@noop {} {\bibfield  {journal} {\bibinfo  {journal} {Journal of Physics
  E: Scientific Instruments}\ }\textbf {\bibinfo {volume} {15}},\ \bibinfo
  {pages} {347} (\bibinfo {year} {1982})}\BibitemShut {NoStop}%
\bibitem [{\citenamefont {Rubin}(2003)}]{rubin2003}%
  \BibitemOpen
  \bibfield  {author} {\bibinfo {author} {\bibnamefont {Rubin}, \bibfnamefont
  {Y.}},\ }\href@noop {} {\emph {\bibinfo {title} {Applied Stochastic
  Hydrogeology}}}\ (\bibinfo  {publisher} {Oxford University Press},\ \bibinfo
  {address} {Oxford},\ \bibinfo {year} {2003})\BibitemShut {NoStop}%
\bibitem [{\citenamefont {Sakho}(2021)}]{sakho2021}%
  \BibitemOpen
  \bibfield  {author} {\bibinfo {author} {\bibnamefont {Sakho}, \bibfnamefont
  {I.}},\ }\href@noop {} {\emph {\bibinfo {title} {Nuclear Physics 1: Nuclear
  Deexcitations, Spontaneous Nuclear Reactions}}}\ (\bibinfo  {publisher} {John
  Wiley \& Sons},\ \bibinfo {year} {2021})\BibitemShut {NoStop}%
\bibitem [{\citenamefont {Severino}(2011)}]{severino2011}%
  \BibitemOpen
  \bibfield  {author} {\bibinfo {author} {\bibnamefont {Severino},
  \bibfnamefont {G.}},\ }\bibfield  {title} {\enquote {\bibinfo {title}
  {Macrodispersion by point-like source flows in randomly heterogeneous porous
  media},}\ }\href {\doibase 10.1007/s11242-011-9758-1} {\bibfield  {journal}
  {\bibinfo  {journal} {Transport in Porous Media}\ }\textbf {\bibinfo {volume}
  {89}},\ \bibinfo {pages} {121--134} (\bibinfo {year} {2011})}\BibitemShut
  {NoStop}%
\bibitem [{\citenamefont {Severino}(2019)}]{severino2019effective}%
  \BibitemOpen
  \bibfield  {author} {\bibinfo {author} {\bibnamefont {Severino},
  \bibfnamefont {G.}},\ }\bibfield  {title} {\enquote {\bibinfo {title}
  {Effective conductivity in steady well-type flows through porous
  formations},}\ }\href {\doibase 10.1007/s00477-018-1639-5} {\bibfield
  {journal} {\bibinfo  {journal} {Stochastic Environmental Research and Risk
  Assessment}\ }\textbf {\bibinfo {volume} {33(3)}},\ \bibinfo {pages}
  {827--835} (\bibinfo {year} {2019})}\BibitemShut {NoStop}%
\bibitem [{\citenamefont {Severino}(2022)}]{severino2022}%
  \BibitemOpen
  \bibfield  {author} {\bibinfo {author} {\bibnamefont {Severino},
  \bibfnamefont {G.}},\ }\bibfield  {title} {\enquote {\bibinfo {title}
  {Dispersion in doublet-type flows through highly anisotropic porous
  formations},}\ }\href@noop {} {\bibfield  {journal} {\bibinfo  {journal}
  {Journal of Fluid Mechanics}\ }\textbf {\bibinfo {volume} {931}},\ \bibinfo
  {pages} {1--13} (\bibinfo {year} {2022})}\BibitemShut {NoStop}%
\bibitem [{\citenamefont {Severino}\ and\ \citenamefont
  {Cuomo}(2020)}]{severino2020uncertainty}%
  \BibitemOpen
  \bibfield  {author} {\bibinfo {author} {\bibnamefont {Severino},
  \bibfnamefont {G.}}\ and\ \bibinfo {author} {\bibnamefont {Cuomo},
  \bibfnamefont {S.}},\ }\bibfield  {title} {\enquote {\bibinfo {title}
  {Uncertainty quantification of unsteady flows generated by line-sources
  through heterogeneous geological formations},}\ }\href@noop {} {\bibfield
  {journal} {\bibinfo  {journal} {SIAM/ASA Journal on Uncertainty
  Quantification}\ }\textbf {\bibinfo {volume} {8}},\ \bibinfo {pages}
  {807--825} (\bibinfo {year} {2020})}\BibitemShut {NoStop}%
\bibitem [{\citenamefont {Severino}, \citenamefont {Leveque},\ and\
  \citenamefont {Toraldo}(2019)}]{severino2019uncertainty}%
  \BibitemOpen
  \bibfield  {author} {\bibinfo {author} {\bibnamefont {Severino},
  \bibfnamefont {G.}}, \bibinfo {author} {\bibnamefont {Leveque}, \bibfnamefont
  {S.}}, \ and\ \bibinfo {author} {\bibnamefont {Toraldo}, \bibfnamefont
  {G.}},\ }\bibfield  {title} {\enquote {\bibinfo {title} {Uncertainty
  quantification of unsteady source flows in heterogeneous porous media},}\
  }\href {\doibase 10.1017/jfm.2019.203} {\bibfield  {journal} {\bibinfo
  {journal} {Journal of Fluid Mechanics}\ }\textbf {\bibinfo {volume} {10}},\
  \bibinfo {pages} {5--26} (\bibinfo {year} {2019})}\BibitemShut {NoStop}%
\bibitem [{\citenamefont {Severino}, \citenamefont {Santini},\ and\
  \citenamefont {Sommella}(2011)}]{severino2011macro}%
  \BibitemOpen
  \bibfield  {author} {\bibinfo {author} {\bibnamefont {Severino},
  \bibfnamefont {G.}}, \bibinfo {author} {\bibnamefont {Santini}, \bibfnamefont
  {A.}}, \ and\ \bibinfo {author} {\bibnamefont {Sommella}, \bibfnamefont
  {A.}},\ }\bibfield  {title} {\enquote {\bibinfo {title} {Macrodispersion by
  diverging radial flows in randomly heterogeneous porous media},}\ }\href
  {\doibase 10.1016/j.jconhyd.2010.12.005} {\bibfield  {journal} {\bibinfo
  {journal} {Journal of contaminant hydrology}\ }\textbf {\bibinfo {volume}
  {123}},\ \bibinfo {pages} {40--49} (\bibinfo {year} {2011})}\BibitemShut
  {NoStop}%
\bibitem [{\citenamefont {Tanaka}\ \emph {et~al.}(2016)\citenamefont {Tanaka},
  \citenamefont {Brunger}, \citenamefont {Campbell}, \citenamefont {Kato},
  \citenamefont {Hoshino},\ and\ \citenamefont {Rau}}]{tanaka2016}%
  \BibitemOpen
  \bibfield  {author} {\bibinfo {author} {\bibnamefont {Tanaka}, \bibfnamefont
  {H.}}, \bibinfo {author} {\bibnamefont {Brunger}, \bibfnamefont {M.~J.}},
  \bibinfo {author} {\bibnamefont {Campbell}, \bibfnamefont {L.}}, \bibinfo
  {author} {\bibnamefont {Kato}, \bibfnamefont {H.}}, \bibinfo {author}
  {\bibnamefont {Hoshino}, \bibfnamefont {M.}}, \ and\ \bibinfo {author}
  {\bibnamefont {Rau}, \bibfnamefont {A.~R.~P.}},\ }\bibfield  {title}
  {\enquote {\bibinfo {title} {Scaled plane-wave born cross sections for atoms
  and molecules},}\ }\href {\doibase 10.1103/RevModPhys.88.025004} {\bibfield
  {journal} {\bibinfo  {journal} {Rev. Mod. Phys.}\ }\textbf {\bibinfo {volume}
  {88}},\ \bibinfo {pages} {025004} (\bibinfo {year} {2016})}\BibitemShut
  {NoStop}%
\bibitem [{\citenamefont {Tartakovsky}, \citenamefont {Tartakovsky},\ and\
  \citenamefont {Meakin}(2008)}]{tartakovsky08}%
  \BibitemOpen
  \bibfield  {author} {\bibinfo {author} {\bibnamefont {Tartakovsky},
  \bibfnamefont {A.~M.}}, \bibinfo {author} {\bibnamefont {Tartakovsky},
  \bibfnamefont {D.~M.}}, \ and\ \bibinfo {author} {\bibnamefont {Meakin},
  \bibfnamefont {P.}},\ }\bibfield  {title} {\enquote {\bibinfo {title}
  {Stochastic~\textsc{L}angevin model for flow and transport in porous
  media},}\ }\href {\doibase 10.1103/PhysRevLett.101.044502} {\bibfield
  {journal} {\bibinfo  {journal} {Phys. Rev. Lett.}\ }\textbf {\bibinfo
  {volume} {101}},\ \bibinfo {pages} {044502} (\bibinfo {year}
  {2008})}\BibitemShut {NoStop}%
\end{thebibliography}%

\end{document}